\documentclass[10pt,preprint2]{aastex}
\usepackage{rotating}	

\newcommand{\percc}{cm$^{-3} $}      %cm-3
\newcommand{\persc}{cm$^{-2} $}      %cm-2
\newcommand{\kks}{K km s$^{-1} $}    %K km s-1
\newcommand{\kms}{km s$^{-1} $}      %km s-1
\newcommand{\HH}{H$_2 $}             %H2
\newcommand{\MOLN}{N$_2 $}           %N2
\newcommand{\AMM}{NH$_{3} $}         %NH3
\newcommand{\HCOP}{HCO$^+ $}         %HCO+
\newcommand{\DCOP}{DCO$^+ $}         %DCO+
\newcommand{\HTCO}{H$_2$CO$ $}       %H2CO
\newcommand{\HTHP}{H$_{3}^{+} $}     %H3+ 
\newcommand{\HTWDP}{H$_{2}$D$^{+} $} %H2D+
\newcommand{\HTHOP}{H$_{3}$O$^{+} $} %H3O+
\newcommand{\CEIO}{C$^{18}$O$ $}     %C18O
\newcommand{\CSEO}{C$^{17}$O$ $}     %C17O
\newcommand{\NTHP}{N$_2$H$^+ $}      % N2H+
\newcommand{\NTDP}{N$_2$D$^+ $}      % N2D+
\newcommand{\DoH}{${\rm {\it N}(N_2D^+)/{\it N}(N_2H^+) }$}

\shorttitle{\NTHP--\NTDP survey in starless cores}
\shortauthors{Crapsi et al.}
\begin{document}

\title{Probing the evolutionary status of starless  cores through \NTHP \ and \NTDP \ observations}

\author{A. Crapsi\altaffilmark{1,2}, P. Caselli\altaffilmark{3}, C.M. Walmsley\altaffilmark{3}, P.C. Myers\altaffilmark{2}, 
M. Tafalla\altaffilmark{4}, C.W. Lee\altaffilmark{5}, T.L. Bourke\altaffilmark{6}} 
%\email{}
\altaffiltext{1}{Universit\`a degli Studi di Firenze, Dipartimento di Astronomia e Scienza dello Spazio, 
                 Largo E. Fermi 5, I-50125 Firenze, Italy}
\altaffiltext{2}{Harvard--Smithsonian Center for Astrophysics, 60 Garden Street, Cambridge, MA 02138, USA}
\altaffiltext{3}{INAF--Osservatorio Astrofisico di Arcetri, Largo E. Fermi 5, I-50125 Firenze, Italy}
\altaffiltext{4}{Observatorio Astron\'omico Nacional (IGN), Alfonso XII, 3, E-28014 Madrid, Spain}
\altaffiltext{5}{Korea Astronomy Observatory, 61-1 Hwaam-dong, Yusung-gu, Daejon 305-348, Korea}
\altaffiltext{6}{Harvard--Smithsonian Center for Astrophysics, Submillimeter Array Project, 645 N.
                 A'ohoku Place, Hilo, HI 96720, USA }

\begin{abstract} 
  We have undertaken a survey of \NTHP \ and \NTDP \ towards  31 low-mass starless cores 
  using the IRAM~30-m telescope. 
  Our main objective has been to determine the abundance ratio of \NTDP \ and \NTHP \ towards 
  the nuclei of these cores and thus to obtain estimates of the degree of deuterium 
  enrichment, a symptom of advanced chemical evolution according to current models.  
  We find that the \DoH \ ratio is larger in more ``centrally concentrated
  cores'' with larger peak \HH \ and \NTHP  \ column density than the sample mean. 
  The deuterium enrichment in starless cores is presently ascribed to depletion of CO 
  in the high density ($>$ 3 $\times$ 10$^4$~\percc)  core nucleus. 
  To substantiate this picture, we compare our results with observations  in dust emission 
  at 1.2~mm and in two transitions of \CEIO . 
  We find a good correlation between deuterium fractionation and $N$(\CEIO)/$N$(\HH)$_{1.2~mm}$ 
  for the nuclei of 14 starless cores.
  We thus identified a set of properties that characterize the most evolved, or ``pre--stellar'', 
  starless cores. These are:  higher \NTHP \ and \NTDP \ column densities, higher \DoH , more 
  pronounced CO depletion, broader \NTHP \ lines with infall asymmetry, higher central \HH \ column  
  densities and a more compact density profile than in the average core. 
  We conclude that this combination of properties gives a reliable indication of
  the evolutionary state of the core.
  Seven cores in our sample (L1521F, OphD, L429, L694, L183, L1544 and TMC2) show the 
  majority of these features and thus are believed to be closer to forming a protostar than
  are the other members of our sample.
  Finally, we note that the subsample of Taurus cores behaves more homogeneously 
  than the total sample, an indication that the external environment could play an 
  important role in the core evolution.
\end{abstract}

\section{Introduction} \label{intro}
  Understanding the physical and kinematic conditions in a molecular cloud core just 
  prior to the onset of gravitational collapse  is of paramount importance if one wishes to study the
  subsequent protostellar evolution. 
  Molecular line measurements can provide the kinematical informations to probe infall motions in the highest
  density nucleus as long as  the observed species is reasonably abundant in the region of interest.
  \\
  An important step forward in starless core studies has been the understanding of the  
  fact that, at temperatures of roughly 10~K and densities above a few times $10^4$~\percc , 
  CO, as well as other carbon-bearing molecules, condenses out onto dust grain surfaces preventing us
  from observing the high-density structures from where one expects the first signs of collapse
  \citep{leger1983}.  This is most clearly seen from the comparison of maps 
  made in the millimeter dust continuum with maps of CO isotopologues 
  \citep{kuiper1996,kramer1999,alves1999,caselli1999}.
  \\  
  For reasons that are thought to have to do with the low binding energy of the 
  N$_2$ molecule on ice mantle surfaces \citep{sadlej1995,bergin1997}, there are 
  some N-containing molecular species that survive in the gas phase at least for densities in 
  the range $10^5$--$10^6$~\percc .
  In particular, maps of \NTHP \ and \AMM \  (as well as their deuterated counterparts) show reasonable 
  agreement with the distribution of dust emission.  Thus, only from observations of
  such species one can hope to probe the  physical and chemical conditions of the most evolved
  pre--stellar cores that  show a  high column density peak in the dust emission.
  \\  
  One can also hope that by studying samples of such high density objects it will be possible to identify 
  the cores where  collapse has already  been initiated or is at the point of doing so. 
  
  The aim of the this paper is to report the results of a survey of 31 cores in  various 
  transitions of \NTHP \ and \NTDP  \ observed with the IRAM~30-m telescope. 

  These observations, besides probing the kinematics of the high density nucleus of starless
  cores, enable us to gauge the deuterium enrichment in our sample.
  This parameter, evaluated from the \NTDP \ over \NTHP \ column density ratio, can be used as
  a tool to discriminate the evolutionary status of a starless core.
  In fact, the disappearance of CO from the gas phase, together with the exothermicity of the production of 
  \HTWDP , leads to  abundances of deuterated species in cold molecular clouds more than 3 orders of magnitudes
  larger than the [D]/[H] elemental abundance \citep[$\simeq 1.5 \times 10^{-5}$ ][]{oliveira2003}.
  The freeze-out of the CO onto the grain mantles decreases the destruction of \HTWDP \ and \HTHP.
  As a result, the enhanced \HTHP \ abundance speeds up the reaction that leads to a higher production of 
  \HTWDP \ and this increases the ratio of  [\HTWDP]/[\HTHP] \citep[e.g.][]{dalgarno1984,roberts2000a}. 
  
  We expect to use the \DoH \ ratio to discriminate the less evolved starless cores from the more advanced 
  pre--stellar cores. For practical purposes in this study, we define ``pre--stellar cores'' to
  be those with an \DoH \ abundance ratio $\ge$ 0.1, and we believe that this approximately corresponds to 
  a core with central density greater than $5 \times 10^5$~\percc \ that is in the
  last phase of its evolution prior to dynamical collapse. We base this partly on the cross-correlation 
  of our \DoH \ measurements with other properties expected to probe an enhanced evolution of the
  starless cores such as  broader line widths, larger $n$(\HH) central density, or larger amount of CO freeze-out.
  For this reason, supplemental dust continuum and \CEIO \ measurements, able to furnish the \HH \ density profile 
  and the  CO depletion, were carried out in cases in which these were not available in the 
  literature.

  In \S~\ref{obs} of this paper, we outline our observational technique both for the \NTHP \ and \NTDP \
  observations and for the dust continuum and \CEIO \ .  In \S~\ref{res}, we
  summarize our direct observational results, and in \S~\ref{ana} we derive physical and chemical quantities 
  for the cores in our sample.  In \S~\ref{dis}, we discuss 
  statistical correlations between the physical properties derived for our sample and
  in \S~\ref{con}, we summarize our conclusions.

\section{Observations} \label{obs}

\subsection{The Sample}
 We have selected 31 cores to be observed in \NTHP \ and \NTDP . 
 The criteria used in the sample selection were not completely homogeneous, but we think that the sample should be 
 representative of evolved starless cores. 

\begin{deluxetable}{lcccccccc}
\tablewidth{0pc}
\tabletypesize{\footnotesize}
\tablecaption{ Core Sample \label{Tsample}}
\tablehead{
   \colhead{core} & \colhead{RA (J2000)} & \colhead{DEC (J2000)} & \colhead{dist} & \colhead{ref} & \colhead{env.} & \colhead{N} & \colhead{1.2-mm} &
   \colhead{\CEIO}       \\
   \colhead{(1)} & \colhead{(2)} & \colhead{(3)} & \colhead{(4)} & \colhead{(5)} & \colhead{(6)} & \colhead{(7)} & \colhead{(8)}  &
   \colhead{(9)}
   }
 \startdata
  L1498      &   04${\rm ^h}$10${\rm ^m}$51\fs5 &   +25\arcdeg09\arcmin58\arcsec  & 140 &  a  &  T  &  25   & y (b)    & y (b) \\
  L1495      &   04${\rm ^h}$14${\rm ^m}$08\fs2 &   +28\arcdeg08\arcmin16\arcsec  & 140 &  b  &  T  &  20   & y (b)    & y (b) \\
  L1495B     &   04${\rm ^h}$18${\rm ^m}$05\fs1 &   +28\arcdeg22\arcmin22\arcsec  & 140 &  a  &  T  &  1    & n        & n     \\
  L1495A-N   &   04${\rm ^h}$18${\rm ^m}$31\fs8 &   +28\arcdeg27\arcmin30\arcsec  & 140 &  a  &  T  &  1    & n        & n     \\
  L1495A-S   &   04${\rm ^h}$18${\rm ^m}$41\fs8 &   +28\arcdeg23\arcmin50\arcsec  & 140 &  a  &  T  &  1    & n        & n     \\
  L1521F     &   04${\rm ^h}$28${\rm ^m}$39\fs8 &   +26\arcdeg51\arcmin35\arcsec  & 140 &  a  &  T  &  89   & y (g)    & y (g) \\
  L1400K     &   04${\rm ^h}$30${\rm ^m}$52\fs1 &   +54\arcdeg51\arcmin55\arcsec  & 140 &  b  &  T  &  20   & n        & n     \\
  L1400A     &   04${\rm ^h}$30${\rm ^m}$56\fs8 &   +54\arcdeg52\arcmin36\arcsec  & 140 &  a  &  T  &  1    & n        & n     \\
  TMC2       &   04${\rm ^h}$32${\rm ^m}$48\fs7 &   +24\arcdeg25\arcmin12\arcsec  & 140 &  a  &  T  &  1    & y (h)    & y (h) \\
  TMC1       &   04${\rm ^h}$41${\rm ^m}$32\fs9 &   +25\arcdeg44\arcmin44\arcsec  & 140 &  a  &  T  &  1    & n        & n     \\
  TMC1C      &   04${\rm ^h}$41${\rm ^m}$38\fs8 &   +25\arcdeg59\arcmin42\arcsec  & 140 &  c  &  T  &  3    & n        & n     \\
  L1507A     &   04${\rm ^h}$42${\rm ^m}$38\fs6 &   +29\arcdeg43\arcmin45\arcsec  & 140 &  a  &  T  &  1    & n        & n     \\
  CB23       &   04${\rm ^h}$43${\rm ^m}$27\fs7 &   +29\arcdeg39\arcmin11\arcsec  & 140 &  a  &  T  &  1    & n        & n     \\
  L1517B     &   04${\rm ^h}$55${\rm ^m}$18\fs8 &   +30\arcdeg38\arcmin04\arcsec  & 140 &  a  &  T  &  21   & y (b)    & y (b) \\
  L1512      &   05${\rm ^h}$04${\rm ^m}$09\fs7 &   +32\arcdeg43\arcmin09\arcsec  & 140 &  a  &  T  &  25   & n        & n     \\
  L1544      &   05${\rm ^h}$04${\rm ^m}$15\fs1 &   +25\arcdeg11\arcmin08\arcsec  & 140 &  a  &  T  &  38   & y (i)    & y (o) \\
  L134A      &   15${\rm ^h}$53${\rm ^m}$33\fs1 &   -04\arcdeg35\arcmin26\arcsec  & 165 &  a  &  B  &  1    & n        & n     \\
  L183       &   15${\rm ^h}$54${\rm ^m}$06\fs4 &   -02\arcdeg52\arcmin23\arcsec  & 165 &  a  &  B  &  77   & y (l)    & y (h) \\
  OphD       &   16${\rm ^h}$28${\rm ^m}$30\fs4 &   -24\arcdeg18\arcmin29\arcsec  & 165 &  d  &  O  &  78   & y (i)    & y (p) \\
  L1689A     &   16${\rm ^h}$32${\rm ^m}$13\fs1 &   -25\arcdeg03\arcmin43\arcsec  & 165 &  e  &  O  &  1    & n        & n     \\
  L1689B     &   16${\rm ^h}$34${\rm ^m}$45\fs8 &   -24\arcdeg37\arcmin50\arcsec  & 165 &  a  &  O  &  1    & n        & n     \\
  L158       &   16${\rm ^h}$47${\rm ^m}$23\fs2 &   -13\arcdeg58\arcmin37\arcsec  & 165 &  a  &  O  &  1    & n        & n     \\
  L234E-S    &   16${\rm ^h}$48${\rm ^m}$08\fs6 &   -10\arcdeg57\arcmin25\arcsec  & 165 &  a  &  O  &  1    & n        & n     \\
  B68        &   17${\rm ^h}$22${\rm ^m}$38\fs9 &   -23\arcdeg49\arcmin46\arcsec  & 125 &  f  &  B  &  1    & y (m)    & y (f) \\
  L492       &   18${\rm ^h}$15${\rm ^m}$46\fs1 &   -03\arcdeg46\arcmin13\arcsec  & 200 &  a  &  A  &  9    & y (h)    & y (h) \\
  L328       &   18${\rm ^h}$17${\rm ^m}$00\fs4 &   -18\arcdeg01\arcmin52\arcsec  & 200 &  d  &  B  &  4    & y (d)    & y (p) \\
  L429       &   18${\rm ^h}$17${\rm ^m}$06\fs4 &   -08\arcdeg14\arcmin00\arcsec  & 200 &  a  &  A  &  68   & y (d)    & y (p) \\
  GF5	     &   18${\rm ^h}$39${\rm ^m}$16\fs4 &   -06\arcdeg38\arcmin15\arcsec  & 200 &  d  &  A  &  1    & n        & n     \\
  L694-2     &   19${\rm ^h}$41${\rm ^m}$04\fs5 &   +10\arcdeg57\arcmin02\arcsec  & 250 &  a  &\nodata&  96 & y (n)    & y (h) \\
  L1197      &   22${\rm ^h}$37${\rm ^m}$02\fs3 &   +58\arcdeg57\arcmin21\arcsec  &\nodata& a &\nodata&  5  & n        & n     \\
  CB246      &   23${\rm ^h}$56${\rm ^m}$49\fs2 &   +58\arcdeg34\arcmin29\arcsec  & 140 &  a  &  B  &  3    & n        & n     \\
\enddata								          

\tablecomments{
Col (1) most common core name; col. (2) and (3) reference position in J2000; col. (4) distance; 
col. (5) references for the position and distance, see below for coding; col. (6) Star forming environment:
T: Taurus molecular cloud; O: Ophiucus molecular cloud; B: Bok globules; A: Aquila rift; 
col. (7) number of observed positions in  both \NTHP \ and \NTDP ; col. (8) availability of a 1.2-mm
continuum map and reference to it, coding below; col. (9) availability of a \CEIO \  map and reference to it, coding below.}
\tablerefs{
a: \citet{lee2001}; 
b: \citet{tafalla2002};  
c: \citet{caselli2002c}; 
d: \citet{bacmann2000}; 
e: \citet{shirley2000}; 
f: \citet{bergin2002};
g: \citet{crapsi2004}; 
h: this work; 
i: \citet{wardthompson1999};  
l: \citet{pagani2003}; 
m: \citet{bianchi2003};
n: Tafalla et al. (in preparation); 
o: \citet{caselli1999}; 
p: \citet{bacmann2002}. }
\end{deluxetable}

``Evolved'' in this context means that there is evidence
 for high density in the form either of  strong \NTHP (1--0) emission 
 ($\gtrsim$~3~\kks \ in T$_{\rm MB}$ scale) or strong 1.2-mm dust continuum 
 ($\gtrsim$~30~mJy/(11\arcsec~beam)) emission or both.   
 The rationale here is that both of these criteria suggest a high central \HH \ column density and hence
 high central pressure, given that \NTHP \ appears to trace the dust well 
 \citep[e.g.][]{tafalla2002}. 

 Of the 31 objects selected in this manner,  24 belong to an earlier  survey of \NTHP \ and CS by \citet{lee2001}, 
 aimed at detecting ``extended'' infall in starless cores through the comparison of the line profiles of optically
 thick (CS) and thin (\NTHP ) tracers. We considered all cores in the \citet{lee2001} sample showing strong 
 \NTHP (1--0) lines and compact morphology. Among these, 13 show infall asymmetry (i.e., CS spectra peaking at lower 
 velocities than \NTHP),  7 show no asymmetry and 4 show outflow asymmetry.

 Of the other 7 cores (L1495, L1400K, TMC1C, L1689A, B68, L328, GF5) in our sample some have been 
 mapped in lines of \AMM \  
 \citep{benson1989}  and \NTHP \  \citep{caselli2002c}  and all have been mapped in dust continuum emission and show
 compact morphology and relatively high central number densities \citep[$n$(\HH) $\ga$ 10$^5$~\percc;  ][]{bacmann2000, 
 shirley2000,tafalla2002}. 

 Finally, the cores in our sample are nearby enough (within 250~pc),  to spatially resolve the dense 
 core nucleus.  The total sample is listed in  Table~\ref{Tsample}; columns (2), (3) and (4) show the
 J2000.0 coordinates and the distance in parsec, taken from the reference in column (5).
 In column (6) we specify if the core is isolated or embedded in a particular star forming region.
 Then in column (7)  we give the total number of positions observed in both \NTHP \ and \NTDP. 
 Finally, in column (8) and (9) we report on the availability of a 1.2-mm map and of a \CEIO \ 
 map with the reference to it.

\subsection{\NTHP \ and \NTDP}
 
   The \NTHP \ and \NTDP \ observations were carried out between April 2002 and January 2003
   using the IRAM~30-m telescope.
   We observed in frequency switch mode and used the autocorrelator as the back end.
   Details on the telescope setting used are given in Table~\ref{Tfreq}. 
\begin{deluxetable}{lccccc}
\tablewidth{0pc}
\tablecaption{ Telescope settings and parameters. \label{Tfreq}}
\tabletypesize{\footnotesize}
\tablehead{
   \colhead{line} & \colhead{frequency} & \colhead{HPBW} & \colhead{F$_{throw}$} & \colhead{T$_{SYS}$} & 
   \colhead{$\Delta v_{res}$}   \\
   \colhead{(1)} & \colhead{(2)} & \colhead{(3)} & \colhead{(4)} & \colhead{(5)} & \colhead{(6)} 
   }
\startdata
\NTHP (1--0)   & 93.1737725 & 26 &  7.5   &  200 & 0.063 \\
\NTHP (3--2)   & 279.511863 &  9 & 14.3   & 2000 & 0.042 \\
\NTDP (2--1)   & 154.217137 & 16 &  7.5   &  400 & 0.038 \\
\NTDP (3--2)   & 231.321966 & 10 & 14.3   &  900 & 0.050 \\
\CEIO (1--0)   & 109.782173 & 22 &\nodata &  170 & 0.026 \\
\CEIO (2--1)   & 219.560357 & 11 &\nodata &  450 & 0.033 \\
\enddata								          
\tablecomments{
Col. (2) line rest frequency~(GHz); 
Col. (3) Half Power Beam Width~(\arcsec)
Col. (4) Freq. Throw~(kHz); 
Col. (5) System Temperature~(K);
Col. (6) Channel Spacing~(\kms)}
\end{deluxetable}
   The frequencies of \NTHP(1--0), \NTDP(2--1) and \NTDP(3--2) were updated following the recent results
   of \citet{dore2004}. A new value for the \NTHP(3--2) frequency was calculated, first evaluating
   the \NTHP \ rotational constant (B $= 46586.87546 \pm 0.00025$~MHz) from the new
   \NTHP(1--0) frequency determination \citep{dore2004}, and     
   then using the centrifugal distortion constant from \citet{verhoeve1990}. 
   In Table~\ref{Tfreq}, we report the frequency of the  $F_1 \, F = 4 \, 5 \rightarrow 3 \, 4$ 
   hyperfine component which has a relative intensity of 17.46\%. 
   The pointing and focus were checked approximately every two hours observing
   planets or quasars in continuum. The pointing corrections never exceeded 4\arcsec .
   The intensity scale was converted to  main beam temperature according to the efficiencies reported
   on the IRAM~30-m Web site.\footnote{Available at  http://www.iram.fr/IRAMES/index.htm}\\
   We first observed the total sample at the peak position given either from previous \NTHP \
   observations \citep{lee2001} or from the dust maps \citep{bacmann2000}.
   Then, we made a map of the cores that showed the strongest \NTDP(2--1) lines:
   L1521F, L1544, TMC2, L183,  OphD, L492, L429 and L694-2.
   TMC2 was mapped only in \NTDP \ and we thus evaluate the ratio of \NTHP \ and \NTDP \ towards the only available
   position observed in \NTHP \  (25\arcsec \ off the \NTDP \ strongest position).
   The \NTHP \ and \NTDP \ maps towards L1495, L1517B, L1498, L1400K and L1512 were taken as part of a different 
   project  on starless cores aimed at determining the ionization degree (P. Caselli et al., in preparation, 
   T. Gatti et al., in preparation); here we just use the  \NTHP \ and \NTDP \ values at the dust peak 
   \citep[determined by][]{tafalla2002}.
   The  L1544 and L1521F data used in the present paper are taken from \citet{caselli2002a} and \citet{crapsi2004}.
   respectively.
   
\subsection{Dust continuum and \CEIO}
   Millimeter continuum  observations, taken with MAMBO (the IRAM~30-m bolometer), were already available
   in the literature for 13 cores in our sample 
  (L1544, OphD: \citealt{wardthompson1999}; 
   L1689B, L328, L429, GF5: \citealt{bacmann2000}; 
   L1498, L1495, L1517B, L1400K: \citealt{tafalla2002}; 
   L183: \citealt{pagani2003};
   L1521F: \citealt{crapsi2004}; 
   L694-2: M. Tafalla et al. 2005, in preparation).  
   Moreover, a 1.2-mm map of B68 taken with SIMBA 
   (the facility bolometer at SEST) was published in \citet{bianchi2003}.
   We obtained a 1.2-mm map of two more cores: TMC2, which presents high \NTDP emission and L492
   which has very strong \NTHP(1--0) lines.
   Observations were carried out by the IRAM staff between January 2003 and February 2003 using the new 
   117-channel bolometer.
   The IRAM~30-m half-power beamwidth (HPBW) at 1.2~mm is approximately 11\arcsec . The calibration error is estimated to be $\sim$10\%.
   \\
   We mapped TMC2 in a 7\arcmin$\times$7\arcmin \ region scanning in the azimuthal
   direction at 5\arcsec s$^{-1}$ with a spacing between the scans of 9\arcsec , reaching a sensitivity of
   6~mJy/beam at the map center.
    The zenith optical depth, measured before and after the map, was 0.2.
   L492 was mapped in a 3\arcmin$\times$3\arcmin \  region scanning in the azimuthal
   direction at 5\arcsec s$^{-1}$ with a spacing between the scans of 8\arcsec; the RMS at map center is
   5~mJy beam$^{-1}$.  The zenith optical depth, measured before taking the map, was 0.2.
   The data reduction (baseline fitting, sky-noise subtraction
   and regridding) was performed using the NIC program of the GAG software developed at IRAM and the Observatoire de 
   Grenoble \citep{broguiere2003}.  
   \\ 
   In the same fashion, we observed the \CEIO (1--0) and (2--1) for those cores where data were not already
   available in literature, i.e., L183, L694-2, L492 and TMC2. We mapped in the on-the-fly mode in a region
   150\arcsec$\times$150\arcsec \ wide (200\arcsec$\times$300\arcsec \ for L183 and 200\arcsec$\times$200\arcsec \ for
   TMC2) scanning in the azimuthal direction.
   Both the separation between two scans and the separation between two dumps were 5\arcsec . 
   \CEIO \ data were already available for L1544 \citep{caselli1999}, OphD, L328, L429 \citep{bacmann2002}, 
   L1495, L1498, L1517B \citep{tafalla2002}, B68 \citep{bergin2002} and L1521F \citep{crapsi2004}.

\section{Results} \label{res}

\subsection{Spectra at dust peak} \label{res1}
\begin{figure}[htbp]
 \begin{center}
 \resizebox{!}{19.5cm}{\includegraphics{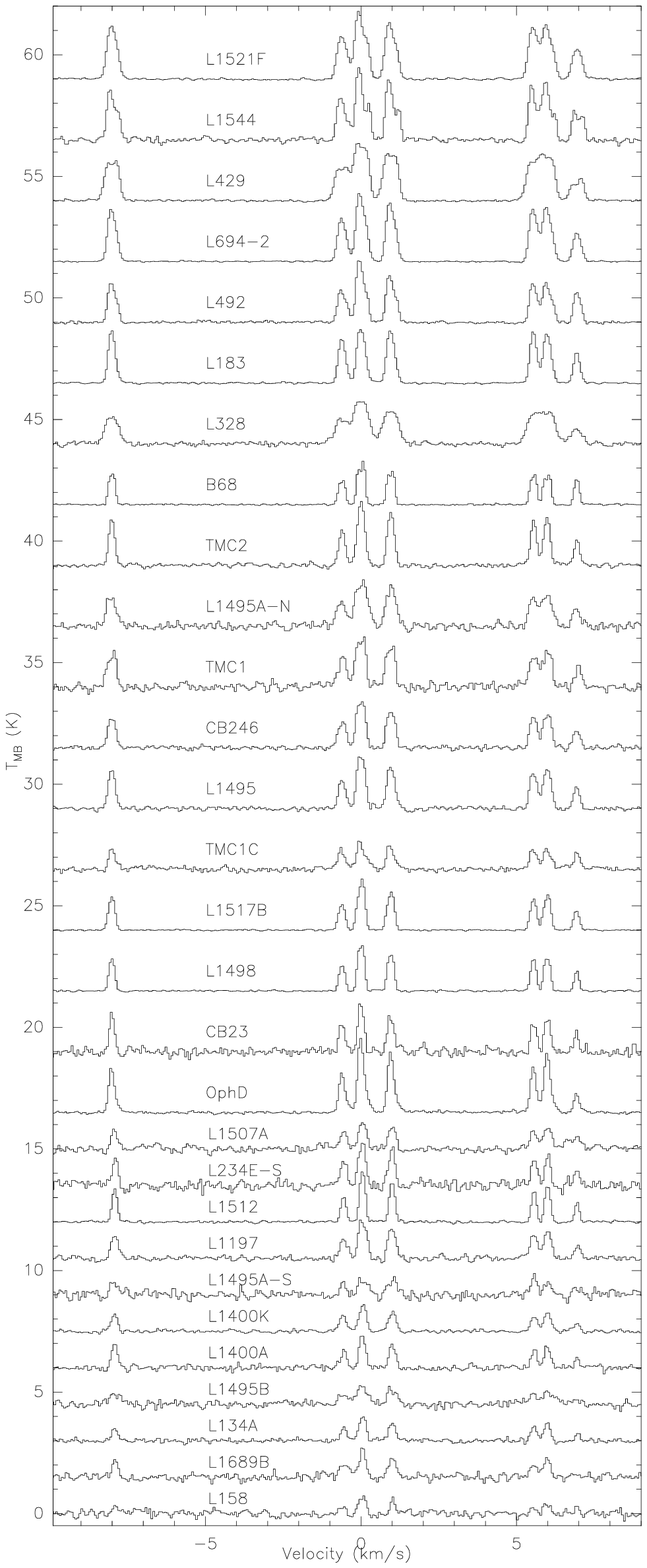}}
 \caption{\NTHP (1--0) spectra at peak sorted by decreasing \NTHP \ column density.
 The two sources with no detection (L1689A and GF5) were omitted. \label{Fspectrah} }
 \end{center}
\end{figure}

\begin{figure}[htbp]
 \begin{center}
 \resizebox{!}{19.1cm}{\includegraphics{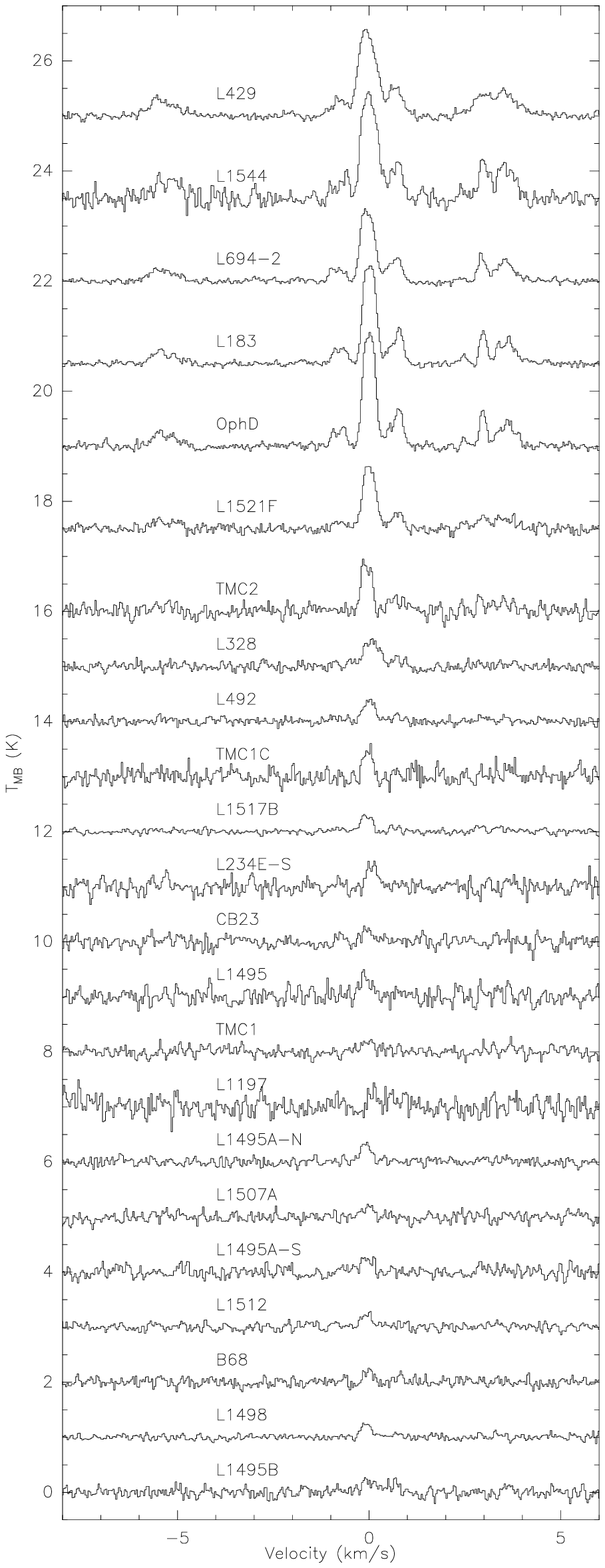}}
 \caption{\NTDP (2--1) spectra at peak sorted by decreasing  \NTDP \ column density. Eight sources with 
 marginal or no detection were omitted  (L1400K, L1400A,  L134A, L1689A, L1689B, L158, 
 GF5, CB246). \label{Fspectrad} }
 \end{center}
\end{figure}

   In Figure~\ref{Fspectrah} and  \ref{Fspectrad}, we show the observed spectra of \NTHP (1--0) and 
   \NTDP (2--1) towards the core dust peaks. The cores where no line was detected were omitted.
   The data were reduced using 
    CLASS, the line analysis program of the GAG software \citep{buisson2002}.   
   Notable in particular is the double-peaked profile with an infall asymmetry observed in L1544,
   as well as the red shoulder or wing seen in several other sources (OphD, L694-2, L492, L1521F, TMC1-C).
   In general, these more ``complex'' profiles are more apparent in the sources where the 
   \NTDP (2--1) line is strong (see below), although there are exceptions (e.g. TMC1 which has a relatively broad
   \NTHP \ line with ``outflow asymmetry'' but weak \NTDP (2--1))\footnote{The TMC1 position observed is 
   close to  the ``CS peak'', see \citet{pratap1997}.}.
   The integrated intensities of these spectra plus the \NTHP (3--2) and \NTDP (3--2) are reported in 
   Table~\ref{Tinten}.
   We also give there the offset of the \NTDP(2--1) peak  from the reference position given in Table~\ref{Tsample}.
   In our sample of 31 cores, we detected \NTHP(1--0) emission with S/N$>$10 in 29 cores. 
   L1689A and GF5, which have a weak detection, will not be considered further in this work.
   We detected \NTDP (2--1) lines with signal-to-noise-ratio (S/N) higher than 10 in 10 cores (L1521F, TMC2, L1517B, L1544, L183, 
   OphD, L492, L328, L429, L694); another 13 were detected with S/N higher than 4.
   In our calculations, as well as the error due to the RMS noise (shown in Table~\ref{Tinten}), 
   we adopt a calibration uncertainty of 10\% for \NTHP (1--0) and 15\% for \NTDP (2--1).

   From hyperfine structure fitting of the spectra in Figure~\ref{Fspectrah} and  \ref{Fspectrad}, we derived the 
   line parameters shown in Table~\ref{Tfit}.
   We report the observed LSR velocities  and line widths in
   the two species and find in first approximation good agreement, suggesting that we
   are sampling the same regions along the line of sight.  
   
   Total optical depth estimates are derived from a simultaneous fit to the hyperfine satellites assuming that hyperfine 
   levels are populated proportional to their statistical weights.  These results show that the total
   optical depth (i.e., the sum of the optical depths of the seven satellites) in \NTHP (1--0) can reach values as high as 20; 
   considering the hyperfine structure of \NTHP (1--0), a total optical depth of 20 means that the main component
   ($F_1 \, F = 2 \, 3 \rightarrow 1 \, 2$), which has a fraction of 7/27 ($\equiv 25.9$\%) of the total line strength, 
   is often optically thick while the  weakest  satellite ($F_1 \, F = 1 \, 0 \rightarrow 1 \, 1$),
   with an intrinsic line strength of 1/27 (or 3.7\%) of the total transition strength, is in almost all cases 
   optically thin (its maximum opacity in our sample is 0.7, towards L183, L1521F and B68). 
   We note, however, that the combination of excitation anomalies 
   \citep[non-LTE effects in the hyperfine level populations, see][]{caselli1995} 
   and highly non-Gaussian  line profiles (e.g., L1544, B68, L492, L1521F, L694-2) 
   increases the uncertainty in the  opacity estimate from the simultaneous fitting of the hyperfine components.
   In fact, the fitting model adopted assumes constant abundance and excitation temperature along 
   the line of sight
   and does not consider variations in velocity and line width.
\begin{deluxetable}{lccccccc}
\tablewidth{0pc}
\tablecaption{Observed line integrated intensities at \NTDP peak.  \label{Tinten} }
\tabletypesize{\footnotesize}
\tablehead{
 \colhead{core}       & \colhead{RA off} & \colhead{DEC off} &   \colhead{I(\NTHP (1--0))} &   \colhead{I(\NTHP (3--2))} &  \colhead{I(\NTDP (2--1))} & 
 \colhead{I(\NTDP (3--2))}  \\ 
            & \colhead{\arcsec}  & \colhead{\arcsec}   &  \colhead{\kks}            &  \colhead{\kks}            &  \colhead{\kks}            &  
	    \colhead{\kks }         
   }
\startdata
   L1498   &  0     &	 0    &  2.27$\pm$0.01 &  0.11$\pm$0.02 &  0.12$\pm$0.02   &	     $<$0.01  \\ 
   L1495   &  40    &	 40   &  3.18$\pm$0.03 &  0.24$\pm$0.04 &  0.23$\pm$0.05   &   0.08$\pm$0.02  \\ 
  L1495B   &  0     &	 0    &  1.35$\pm$0.05 &        $<$0.09 &  0.14$\pm$0.04   &   0.30$\pm$0.09  \\ 
L1495A-N   &  0     &	 0    &  3.46$\pm$0.05 &        $<$0.09 &  0.18$\pm$0.02   &	     $<$0.09  \\ 
L1495A-S   &  0     &	 0    &  1.08$\pm$0.06 &        $<$0.12 &  0.15$\pm$0.04   &	     $<$0.11  \\ 
  L1521F   & -10    &	 0    &  5.86$\pm$0.02 &  0.74$\pm$0.04 &  0.97$\pm$0.02   &   0.29$\pm$0.01  \\ 
  L1400K   & -20    &	-40   &  1.39$\pm$0.02 &  0.10$\pm$0.04 &      $<$0.04     &	     $<$0.02  \\ 
  L1400A   &  0     &	 0    &  1.48$\pm$0.04 &  0.25$\pm$0.12 &  0.08$\pm$0.02   &   0.13$\pm$0.07  \\ 
    TMC2   &  0     &	 40   &  3.48$\pm$0.03 &  0.32$\pm$0.11 &  0.66$\pm$0.04   &   0.17$\pm$0.05  \\ 
    TMC1   &  0     &	 0    &  3.51$\pm$0.06 &  0.35$\pm$0.11 &  0.20$\pm$0.03   &	     $<$0.10  \\ 
   TMC1C   &  0     &	 40   &  1.87$\pm$0.03 &        $<$0.19 &  0.24$\pm$0.04   &   0.08$\pm$0.06  \\ 
  L1507A   &  0     &	 0    &  1.78$\pm$0.05 &  0.21$\pm$0.12 &  0.15$\pm$0.03   &   0.15$\pm$0.07  \\ 
    CB23   &  0     &	 0    &  2.22$\pm$0.06 &  0.20$\pm$0.14 &  0.21$\pm$0.04   &   0.37$\pm$0.10  \\ 
  L1517B   & -15    &	-15   &  2.57$\pm$0.01 &  0.11$\pm$0.02 &  0.26$\pm$0.02   &	     $<$0.01  \\ 
   L1512   &  0     &	 0    &  2.11$\pm$0.02 &        $<$0.05 &  0.17$\pm$0.03   &	     $<$0.02  \\ 
   L1544   &  20    &	-20   &  5.46$\pm$0.04 &  0.56$\pm$0.09 &  2.23$\pm$0.05   &   0.55$\pm$0.03  \\ 
   L134A   &  0     &	 0    &  1.13$\pm$0.03 &        $<$0.25 &      $<$0.04     &	     $<$0.07  \\ 
    L183   &  30    &	 0    &  4.20$\pm$0.01 &  0.56$\pm$0.04 &  1.63$\pm$0.02   &   0.48$\pm$0.01  \\ 
    OphD   & -20    &	-50   &  3.71$\pm$0.02 &  0.39$\pm$0.05 &  1.75$\pm$0.02   &   0.59$\pm$0.02  \\ 
  L1689A   &  0     &	 0    &  0.15$\pm$0.07 &        $<$0.13 &      $<$0.05     &	     $<$0.06  \\ 
  L1689B   &  0     &	 0    &  1.27$\pm$0.05 &        $<$0.13 &  0.09$\pm$0.04   &   0.12$\pm$0.06  \\ 
    L158   &  0     &	 0    &  0.61$\pm$0.05 &        $<$0.11 &      $<$0.04     &	     $<$0.05  \\ 
 L234E-S   &  0     &	 0    &  2.06$\pm$0.07 &  0.40$\pm$0.26 &  0.26$\pm$0.04   &	     $<$0.32  \\ 
     B68   &  0     &	 0    &  2.37$\pm$0.01 &  0.17$\pm$0.02 &  0.13$\pm$0.03   &         $<$0.03  \\ 
    L492   &  20    &	 20   &  3.95$\pm$0.02 &        $<$0.10 &  0.30$\pm$0.02   &         $<$0.04  \\ 
    L328   &  0     &	 0    &  4.46$\pm$0.03 &  0.42$\pm$0.09 &  0.38$\pm$0.03   &   0.19$\pm$0.04  \\ 
    L429   & -20    &	 20   &  6.11$\pm$0.02 &  0.57$\pm$0.05 &  2.03$\pm$0.02   &   0.62$\pm$0.02  \\ 
     GF5   &  0     &	 0    &  0.39$\pm$0.05 &        $<$0.07 &      $<$0.04     &         $<$0.10  \\ 
  L694-2   &  0     &	 0    &  5.13$\pm$0.01 &  0.36$\pm$0.02 &  1.35$\pm$0.02   &   0.37$\pm$0.01  \\ 
   L1197   &  0     &	 0    &  2.15$\pm$0.04 &        $<$0.16 &  0.18$\pm$0.06   &	     $<$0.05  \\ 
   CB246   & -60    &	-20   &  2.91$\pm$0.03 &        $<$0.07 &  0.11$\pm$0.04   &   0.22$\pm$0.06  \\ 
\enddata								          
\tablecomments{Offsets are referred to the positions in Table~\ref{Tsample}. Temperatures are in main beam temperature scale.
Calibration errors were not included in the error estimate. Upper limits assume the line-width of \NTHP (1--0) where available or 
0.17~\kms \, the smallest line-width observed in our sample. }
\end{deluxetable}
\begin{deluxetable}{l|cccc|cccc}
\tablewidth{0pc}
\tablecaption{Results of the hyperfine structure fitting on the \NTHP(1--0) and \NTDP(2--1) spectra at the peak position of each core. 
\label{Tfit}}
\tabletypesize{\footnotesize}
\tablehead{
            &            \multicolumn{4}{c|}{\NTHP(1--0)}          &	                  \multicolumn{4}{c}{\NTDP(2--1)}	           \\
 core      & V$_{LSR}$         &   $\Delta V$    &  $\tau$       &  T$_{EX}$     &    V$_{LSR}$	        & $\Delta V$      &  $\tau$      &  T$_{EX}$     \\
           &  \kms             &  \kms           &               &   K           &     \kms             &  \kms           &	         &      K     
}
\startdata
 L1498      &  7.822$\pm$0.001 & 0.185$\pm$0.001 &  11.5$\pm$0.3  & 4.5$\pm$0.1   &    7.778$\pm$0.010  & 0.176$\pm$0.029  & 0.1	  &  4.5         \\
 L1495      &  6.807$\pm$0.001 & 0.237$\pm$0.003 &   9.6$\pm$0.6  & 4.8$\pm$0.2   &    6.755$\pm$0.025  & 0.290$\pm$0.061  & 0.1	  &  4.8         \\
 L1495B     &  6.633$\pm$0.008 & 0.390$\pm$0.018 &   0.1	  &   4.8           &    6.683$\pm$0.065  & 0.500$\pm$0.216  & 0.1	  &  4.8         \\
 L1495A-N   &  7.296$\pm$0.003 & 0.327$\pm$0.007 &   7.6$\pm$0.9  & 4.7$\pm$0.3   &    7.237$\pm$0.013  & 0.244$\pm$0.035  & 0.1	  &  4.7         \\
 L1495A-S   &  7.294$\pm$0.007 & 0.214$\pm$0.015 &  21.3$\pm$5.3  & 4.4$\pm$0.4   &    7.264$\pm$0.024  & 0.284$\pm$0.059  & 0.1	  &  4.4         \\
 L1521F     &  6.472$\pm$0.001 & 0.299$\pm$0.001 &  17.9$\pm$0.3  & 4.9$\pm$0.1   &    6.505$\pm$0.004  & 0.268$\pm$0.010  & 2.2$\pm$0.4  &  4.6$\pm$0.3 \\
 L1400K     &  3.196$\pm$0.002 & 0.231$\pm$0.005 &   6.5$\pm$0.8  & 4.4$\pm$0.5   &	\nodata		&  \nodata  	   &  \nodata	  &  \nodata	 \\
 L1400A     &  3.355$\pm$0.002 & 0.191$\pm$0.006 &   7.7$\pm$1.2  & 4.4$\pm$0.6   &    3.250$\pm$0.036  & 0.175$\pm$0.061  & 0.1	  &  4.4	 \\
 TMC2	    &  6.193$\pm$0.001 & 0.210$\pm$0.002 &   8.7$\pm$0.4  & 5.1$\pm$0.2   &    6.157$\pm$0.006  & 0.167$\pm$0.015  & 4.5$\pm$1.1  &  5.0$\pm$0.3 \\
 TMC1	    &  5.856$\pm$0.003 & 0.269$\pm$0.007 &  11.1$\pm$1.2  & 4.6$\pm$0.3   &    5.870$\pm$0.040  & 0.409$\pm$0.113  & 0.1	  &  4.6	 \\
 TMC1C      &  5.196$\pm$0.003 & 0.212$\pm$0.006 &  18.4$\pm$2.2  & 4.4$\pm$0.5   &    5.193$\pm$0.009  & 0.132$\pm$0.017  & 0.1	  &  4.4	 \\
 L1507A     &  6.163$\pm$0.004 & 0.220$\pm$0.009 &  12.2$\pm$2.1  & 4.4$\pm$0.6   &    6.159$\pm$0.030  & 0.254$\pm$0.071  & 0.1	  &  4.4	 \\
 CB23	    &  6.015$\pm$0.002 & 0.166$\pm$0.006 &  14.6$\pm$2.2  & 4.4$\pm$0.3   &    5.985$\pm$0.027  & 0.233$\pm$0.075  & 0.1	  &  4.4	 \\
 L1517B     &  5.835$\pm$0.001 & 0.215$\pm$0.001 &   8.8$\pm$0.2  & 4.7$\pm$0.1   &    5.796$\pm$0.008  & 0.218$\pm$0.017  & 0.1	  &  4.7	 \\
 L1512      &  7.121$\pm$0.001 & 0.174$\pm$0.001 &   7.3$\pm$0.4  & 5.0$\pm$0.2   &    7.091$\pm$0.018  & 0.274$\pm$0.076  & 0.1	  &  4.8	 \\
 L1544      &  7.143$\pm$0.002 & 0.315$\pm$0.004 &  12.6$\pm$0.7  & 5.0$\pm$0.2   &    7.181$\pm$0.004  & 0.289$\pm$0.008  & 5.2$\pm$0.4  &  4.8$\pm$0.2 \\
 L134A      &  2.665$\pm$0.002 & 0.215$\pm$0.007 &   4.3$\pm$1.0  & 4.4$\pm$0.6   &	\nodata		&  \nodata  	   &  \nodata	  &  \nodata	 \\
 L183	    &  2.413$\pm$0.001 & 0.211$\pm$0.001 &  20.3$\pm$0.2  & 4.8$\pm$0.1   &    2.459$\pm$0.001  & 0.214$\pm$0.002  & 4.7$\pm$0.2  &  4.9$\pm$0.1 \\
 OphD	    &  3.478$\pm$0.001 & 0.223$\pm$0.001 &   4.4$\pm$0.1  & 7.1$\pm$0.2   &    3.515$\pm$0.001  & 0.204$\pm$0.003  & 4.3$\pm$0.2  &  5.3$\pm$0.2 \\
 L158	    &  3.942$\pm$0.007 & 0.229$\pm$0.014 &   0.1	  &   4.8         &	\nodata		&  \nodata  	   &  \nodata	  &   \nodata 	 \\
 L1689B     &  3.481$\pm$0.005 & 0.262$\pm$0.011 &   0.1	  &   4.8         &    3.642$\pm$0.058  & 0.373$\pm$0.148  & 0.1	  &  4.8	 \\
 L234E-S    &  3.164$\pm$0.003 & 0.198$\pm$0.008 &   6.9$\pm$1.5  & 4.7$\pm$0.5   &    3.286$\pm$0.015  & 0.182$\pm$0.055  & 0.1	  &  4.7	 \\
 B68        &  3.364$\pm$0.001 & 0.176$\pm$0.001 &  26.2$\pm$1.0  & 4.5$\pm$0.5   &    3.372$\pm$0.016  & 0.124$\pm$0.036  & 0.1          &  4.5         \\  
 L492       &  7.701$\pm$0.001 & 0.263$\pm$0.001 &  11.8$\pm$0.2  & 4.8$\pm$0.1   &    7.726$\pm$0.008  & 0.222$\pm$0.022  & 2.6$\pm$0.8  &  4.4$\pm$0.3 \\ 
 L328	    &  6.707$\pm$0.002 & 0.438$\pm$0.005 &   8.1$\pm$0.5  & 4.6$\pm$0.2   &    6.797$\pm$0.012  & 0.419$\pm$0.028  & 0.1	  &  4.6	 \\
 L429       &  6.719$\pm$0.001 & 0.394$\pm$0.001 &  12.3$\pm$0.1  & 4.9$\pm$0.1   &    6.705$\pm$0.002  & 0.360$\pm$0.005  & 4.3$\pm$0.2  &  4.4$\pm$0.1 \\
 L694-2     &  9.574$\pm$0.001 & 0.266$\pm$0.001 &  13.9$\pm$0.1  & 5.2$\pm$0.1   &    9.567$\pm$0.002  & 0.244$\pm$0.004  & 5.5$\pm$0.2  &  4.1$\pm$0.1 \\
 L1197      & -3.147$\pm$0.002 & 0.249$\pm$0.006 &   6.5$\pm$0.9  & 4.5$\pm$0.3   &   -3.020$\pm$0.016  & 0.180$\pm$0.093  & 0.1	  &  4.5         \\
 CB246      & -0.830$\pm$0.001 & 0.245$\pm$0.003 &  10.3$\pm$0.7  & 4.5$\pm$0.2   &   -0.730$\pm$0.025  & 0.126$\pm$0.039  & 0.1	  &  4.5         \\

\enddata								          
\tablecomments{The hyperfine fitting model assumes constant T$_{EX}$ and homogenous abundance and do not consider 
velocity gradients effects along the line of sight. L1689A and GF5 spectra were not fitted because of poor S/N. 
}
\end{deluxetable}

\begin{deluxetable}{l|ccc|ccc}
\tablewidth{0pc}
\tabletypesize{\footnotesize}
\tablecaption{Results of the hyperfine structure fitting on the \NTHP(3--2) and \NTDP(3--2) spectra at the peak position of each core.
Only the fits to high S/N ($>5$) spectra are reported. \label{Tfit1mm}}
\tablehead{
            &            \multicolumn{3}{c|}{\NTHP(3--2)}               &	    \multicolumn{3}{c}{\NTDP(3--2)}	   \\
 core       & V$_{LSR}$        & $\Delta V$	   &  $\tau$	  & V$_{LSR}$	      & $\Delta V$	   &  $\tau$	   \\
            &  \kms            &  \kms  	   &		  &  \kms	      &  \kms		   &
 }
\startdata
 L1498     &   7.786$\pm$0.015  &   0.169$\pm$0.040  &      $<$0.1    &    \nodata	    &	   \nodata	 &    \nodata      \\
 L1495     &   6.811$\pm$0.030  &   0.288$\pm$0.052  &	    $<$0.1    &    \nodata	    &	   \nodata	 &    \nodata 	   \\
 L1521F    &   6.391$\pm$0.010  &   0.172$\pm$0.021  &   6.2$\pm$1.5  &   6.507$\pm$0.006   &  0.222$\pm$0.025   &  1.2$\pm$0.9    \\
 L1517B    &   5.816$\pm$0.030  &   0.219$\pm$0.130  &	    $<$0.2    &    \nodata	    &	   \nodata	 &    \nodata 	   \\
 L1544     &   7.138$\pm$0.018  &   0.193$\pm$0.080  &	    $<$0.1    &   7.208$\pm$0.012   &  0.342$\pm$0.026   &     $<$0.1 	   \\
 L183      &   2.412$\pm$0.012  &   0.220$\pm$0.036  &   3.5$\pm$1.6  &   2.445$\pm$0.003   &  0.205$\pm$0.012   &  1.2$\pm$0.4    \\
 OphD      &   3.466$\pm$0.011  &   0.131$\pm$0.019  &   6.7$\pm$2.2  &   3.503$\pm$0.004   &  0.196$\pm$0.013   &  2.6$\pm$0.6    \\
 B68       &   3.370$\pm$0.017  &   0.231$\pm$0.022  &   0.8$\pm$0.4  &    \nodata	    &	   \nodata	 &    \nodata 	   \\
 L429      &   6.673$\pm$0.029  &   0.373$\pm$0.037  &   2.7$\pm$0.4  &   6.696$\pm$0.009   &  0.262$\pm$0.020   &  5.2$\pm$0.9    \\
 L694-2    &   9.532$\pm$0.014  &   0.308$\pm$0.060  &   1.7$\pm$1.4  &   9.560$\pm$0.005   &  0.242$\pm$0.016   &  1.5$\pm$0.5    \\
\enddata								          
\end{deluxetable}	
\begin{deluxetable}{l|ccc|ccc}
\tablewidth{0pc}
\tabletypesize{\footnotesize}
\tablecaption{Integrated intensities and results of the gaussian fitting of 
the \CEIO(1--0) and (2--1) spectra towards the peak position for the four cores newly observed.
\label{TfitCO}}
\tablehead{
            &  \multicolumn{3}{c|}{\CEIO(1--0)} & \multicolumn{3}{c}{\CEIO(2--1)}      \\
 core       & I(\CEIO(1--0)) & V$_{LSR}$       &  $\Delta V$     & I(\CEIO(2--1))  & V$_{LSR}$	     & $\Delta V$    \\
            & K \kms         & \kms            &   \kms	         & K \kms          &  \kms  	     &  \kms  	       }
\startdata
 L1521F     & 1.80$\pm$0.06  & 6.458$\pm$0.006 & 0.438$\pm$0.014 &  1.80$\pm$0.15  & 6.427$\pm$0.015 & 0.447$\pm$0.034    \\
 TMC2       & 1.05$\pm$0.04  & 6.313$\pm$0.007 & 0.401$\pm$0.014 &  1.09$\pm$0.07  & 6.229$\pm$0.010 & 0.392$\pm$0.023    \\
 L183       & 1.89$\pm$0.04  & 2.518$\pm$0.006 & 0.545$\pm$0.016 &  2.14$\pm$0.06  & 2.556$\pm$0.008 & 0.579$\pm$0.019    \\
 L492	    & 0.76$\pm$0.04  & 7.786$\pm$0.017 & 0.651$\pm$0.042 &  0.58$\pm$0.05  & 7.786$\pm$0.022 & 0.529$\pm$0.055    \\
 L694-2     & 1.21$\pm$0.04  & 9.574$\pm$0.007 & 0.433$\pm$0.015 &  1.01$\pm$0.04  & 9.574$\pm$0.006 & 0.415$\pm$0.015    \\
\enddata								          
\end{deluxetable}	

\begin{figure}[htbp]
 \begin{center}
  \resizebox{8cm}{!}{\includegraphics{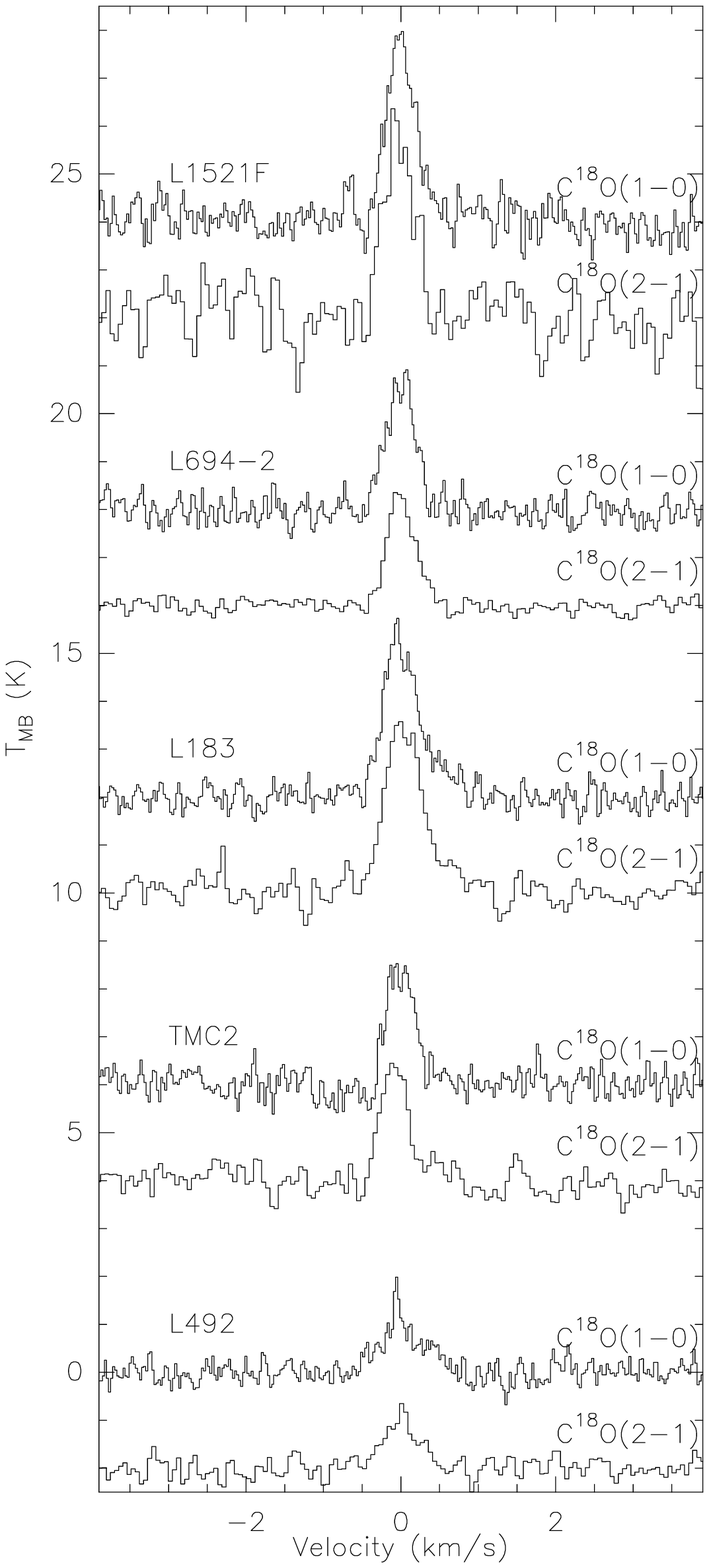}}
 \caption{\CEIO(1--0) and (2--1) spectra towards L1521F, L694-2, L183, TMC2 and L492. The shown spectra 
 are the arithmetic means of all the observations in the 10\arcsec \ strip centered around the peak 
 position and going parallel to the Right Ascension direction. \label{FspectraCO} }
 \end{center}
\end{figure}

   To crudely estimate the variations on the more complex spectra, we derived the opacity of the 
   \NTHP(1--0) spectra in B68 and L1544 by the fit of only a  couple of hyperfines and repeating 
   that for all of the different couples.
   In this way, we found a scatter of a factor of 2 in the resulting opacities around the value derived by the simultaneous 
   fit of all  seven components.    
   The results of the hyperfine structure fitting of the high-S/N ($>$ 5) 1mm spectra are reported in Table~\ref{Tfit1mm}.
   We also show in Figure~\ref{FspectraCO} the \CEIO(1--0) and (2--1) spectra towards the peak of L1521F, L694-2, 
   L183, TMC2 and L492.
   Integrated intensities and results of the Gaussian fits for these observations are reported in Table~\ref{TfitCO}.

\subsection{Maps of individual sources}

  A detailed description of the maps obtained in this work can be found in 
  the Appendix~\ref{maps}, where we also analyse core shapes and kinematics (Appendix~\ref{shape}) 
  of individual objects.  Here we briefly summarize the main results.  Our new 
  maps confirm the well known finding that \NTHP \ and \NTDP \ well trace the 
  millimeter dust continuum emission, unlike CO (e.g. Caselli et al. 1999; 
  Bergin et al. 2001; Tafalla et al. 2002). We also find that the majority of the
  cores are ``cometary'' in shape, suggesting that the cores have been 
  either compressed by external agents or have moved relatively to the 
  surrounding environment (but see Walsh \& Myers 2004). In general, they do not show an axisymmetric 
  morphology. Finally, we studied the kinematics in individual cores, evaluating
  the ``total'' and ``local'' velocity gradients and  finding
  that only in one case (L694-2) the \NTHP (1--0) line width increases toward 
  the core center, indicating the presence of enhanced dynamical activity 
  (likely due to infall), as already found in L1544 (Caselli et al. 2002a) and
  L1521F (Crapsi et al. 2004).  

  In the rest of the paper we concentrate on the physical and chemical parameters at the relative peak 
  of the whole source sample.

\section{Analysis} \label{ana}
  In this section we derive the physical and chemical properties that can best discriminate the evolutionary status 
  of a starless core, either in a chemical sense, as a more pronounced deuterium enrichment and a higher  
  CO depletion factor, or  for dynamical reasons, like the presence of a denser and more centrally concentrated \HH \
  distribution or broader \NTHP \ line widths. We expect to characterize through these parameters the pre--stellar
  cores, i.e., the subset of starless cores that present enhanced chemical evolution and/or dynamical activity.

 \subsection{Column densities of \NTHP \ and \NTDP \ and the deuterium fractionation} \label{dfr}

  The deuterium fractionation can be evaluated from observations of a hydrogen bearing molecule and one of its 
  deuterated counterparts by calculating the ratio of their column densities. Recent work 
  has made use of \HCOP, \HTCO, HNC, \NTHP \ and \AMM, as well as their single or multiple  deuterated isotopologues, to 
  evaluate this parameter in starless cores 
  \citep{caselli1998,roueff2000,tine2000,bacmann2003,caselli2002b,hirota2003,crapsi2004}. 
  One should note, however, given the depletion of \HCOP \ and \HTCO \ in the core nucleus 
  \citep{caselli2002b,carey1998}, that the deuterium  fractionation estimated from this species is 
  relative to an outer shell and not to the high-density nucleus. 

\begin{figure}[!tbp]
 \begin{center}
 \resizebox{8cm}{!}{\includegraphics{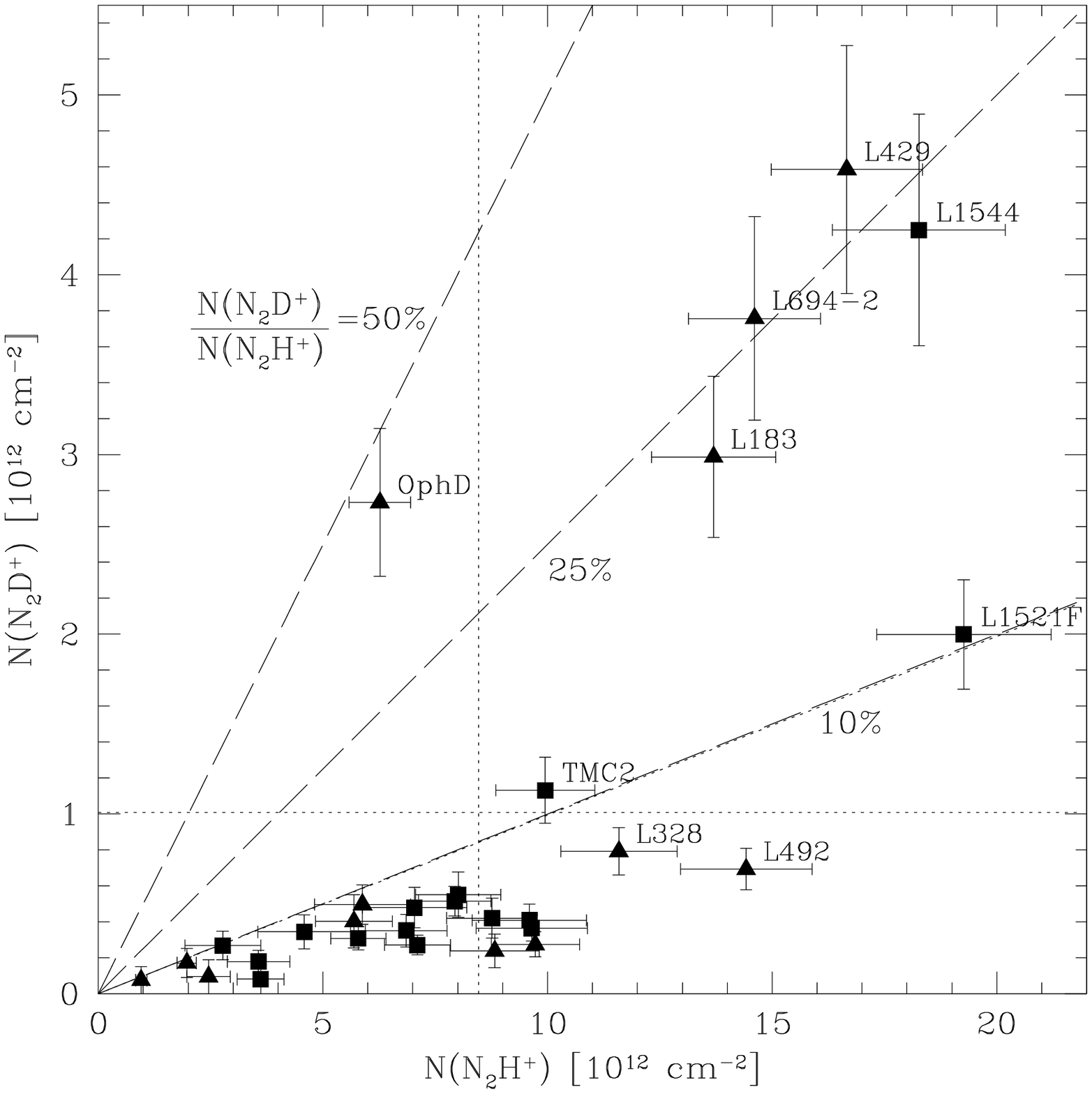}}
 \caption{\NTHP \ column density vs. \NTDP \ column density at \NTDP \ peak position. 
 Deuterium fractionation can be read from the radial dashed lines (corresponding to \DoH\ of 0.1, 0.25 and 0.5). 
 Dotted lines are the arithmetic means of the distributions. 
 Cores belonging to the Taurus Molecular Cloud are marked by a square. \label{Fn2dn2h}}
 \end{center}
\end{figure}
  
  In this work we derived \NTHP \ and \NTDP \ column densities assuming  a constant excitation 
  temperature \citep[the CTEX method: see the appendix of ][]{caselli2002b}. The excitation temperature of \NTHP(1--0) 
  was derived from hyperfine fitting of the spectra with well constrained opacities 
  ($\tau/\sigma_{\tau}>3$, where the total optical depth $\tau$ and its associated
  error $\sigma_{\tau}$ have been derived assuming a constant excitation
  temperature for the hyperfine components; see also Table~\ref{Tfit}); 
  otherwise the spectrum was assumed optically thin and the excitation temperature 
  was set to 4.8~K (the average excitation temperature in our sample).
  The excitation temperature of \NTDP(2--1) was set equal to that of  \NTHP (1--0) in
  those cases in which the opacity was not well constrained. The effect of underestimating the \NTDP(2--1) 
  excitation temperature by,  e.g., 1~K would be a decrease of $N$(\NTDP) by 30\%.\\
  As noted above, in the case of \NTHP , the optically thin assumption is
  often incorrect and hence we have used the intensity of the ($F_1 \, F = 1 \, 0 \rightarrow 1 \, 1$)
  (weak) component normalized appropriately in order to estimate the J=0 column density.

  The final result can be seen in Figure~\ref{Fn2dn2h}, where the \NTHP \ column density is plotted versus the
  \NTDP \ column density.
  The estimated \DoH \ ratio varies between 0.44 in Oph D and an upper limit of 0.02 in L1400K.
  It is noticeable that with the exception of OphD, all cores with \DoH \ ratio above 0.10 have 
  \NTHP \ column densities above $10^{13}$~\persc , whereas cores with  \DoH \ lower than 0.10 have
  $N$(\NTHP) lower than $10^{13}$~\persc \  (with a few exceptions). Thus, the impression from Figure~\ref{Fn2dn2h} is that
  there is a trend in the sense that the \DoH \ ratio correlates positively  with \NTHP \ column density.  
  Table~\ref{Tcolden} summarize the above numbers.
\begin{deluxetable}{lccccccc}
\tablewidth{0pt}
\tablecaption{\NTHP \ and \NTDP \ column densities and relative ratio towards the \NTDP (2--1) 
integrated intensity peak.
 \label{Tcolden} }
\tabletypesize{\footnotesize}
\tablehead{
 core {~~~~~~}      & {~~~}$N$(\NTHP){~~~}        &  {~~~} $N$(\NTDP)   {~~~}     & {~~} \DoH   {~~}           \\  
            & $10^{12}~$\persc & $10^{11}~$\persc                           }
\startdata
       L1498    &     7.1$\pm$0.7   &	   2.7$\pm$0.6    &    0.04$\pm$0.01  \\
       L1495    &     8.8$\pm$1.0   &	   4.2$\pm$1.1    &    0.05$\pm$0.01  \\
      L1495B    &     2.8$\pm$0.8   &	   2.7$\pm$0.8    &    0.10$\pm$0.04  \\
    L1495A-N    &     9.6$\pm$1.2   &	   3.6$\pm$0.7    &    0.04$\pm$0.01  \\
    L1495A-S    &     4.6$\pm$1.0   &	   3.4$\pm$1.0    &    0.08$\pm$0.03  \\
      L1521F    &    19.3$\pm$1.9   &	  20.0$\pm$3.0    &    0.10$\pm$0.02  \\
      L1400K    &     3.6$\pm$0.5   &	        $<$0.8    &          $<$0.02  \\
      L1400A    &     3.6$\pm$0.7   &	   1.8$\pm$0.6    &    0.05$\pm$0.02  \\
        TMC2    &     9.9$\pm$1.1   &	  11.3$\pm$1.8    &    0.11$\pm$0.02  \\
        TMC1    &     9.6$\pm$1.3   &	   4.1$\pm$0.9    &    0.04$\pm$0.01  \\
       TMC1C    &     8.0$\pm$0.9   &	   5.5$\pm$1.3    &    0.07$\pm$0.02  \\
      L1507A    &     6.9$\pm$0.9   &	   3.5$\pm$0.9    &    0.05$\pm$0.01  \\
        CB23    &     7.0$\pm$1.2   &	   4.8$\pm$1.1    &    0.07$\pm$0.02  \\
      L1517B    &     7.9$\pm$0.8   &	   5.1$\pm$0.8    &    0.06$\pm$0.01  \\
       L1512    &     5.8$\pm$0.6   &	   3.1$\pm$0.6    &    0.05$\pm$0.01  \\ 
L1544\tablenotemark{a}    &    18.3$\pm$1.9   &	  42.5$\pm$6.4    &    0.23$\pm$0.04  \\
       L134A    &     2.5$\pm$0.5   &	        $<$0.9    &          $<$0.04  \\
        L183    &    13.7$\pm$1.4   &	  29.9$\pm$4.5    &    0.22$\pm$0.04  \\
        OphD    &     6.3$\pm$0.7   &	  27.3$\pm$4.1    &    0.44$\pm$0.08  \\
        L158    &     1.0$\pm$0.1   &	        $<$0.7    &          $<$0.08  \\
      L1689B    &     2.0$\pm$0.2   &	   1.7$\pm$0.8    &    0.09$\pm$0.04  \\
     L234E-S    &     5.9$\pm$1.1   &	   5.0$\pm$1.1    &    0.08$\pm$0.02  \\
         B68    &     9.7$\pm$1.0   &      2.7$\pm$0.7    &    0.03$\pm$0.01  \\
        L492    &    14.4$\pm$1.5   &	   6.9$\pm$1.1    &    0.05$\pm$0.01  \\
        L328    &    11.6$\pm$1.3   &	   7.9$\pm$1.3    &    0.07$\pm$0.01  \\
        L429    &    16.7$\pm$1.7   &	  45.9$\pm$6.9    &    0.28$\pm$0.05  \\
      L694-2    &    14.6$\pm$1.5   &	  37.6$\pm$5.7    &    0.26$\pm$0.05  \\
       L1197    &     5.7$\pm$0.9   &	   4.0$\pm$1.5    &    0.07$\pm$0.03  \\
       CB246    &     8.8$\pm$1.0   &	   2.4$\pm$0.9    &    0.03$\pm$0.01  \\
       
\enddata								          
\tablenotetext{a}{The difference in the L1544 \DoH \ given by \citet{caselli2002b} (0.26) and  
  here (0.23) is due to the different approximation used. They evaluated the column density from 
  all the seven hyperfine components correcting for the high optical depth (see their eq. A1) while here we 
  used the optically thin approximation applied on the weakest component only.}   
       
\end{deluxetable}

  We also determined the \NTHP \ and \NTDP \ column densities in the Large Velocity Gradient (LVG) approximation
  in those cores with high-S/N \NTHP \ and \NTDP(3--2) spectra (see Section~\ref{LVG}).
  Results are shown in Table \ref{TLVG}.
  The LVG column density determinations are found to agree with the CTEX values with a dispersion of 30\%.

  \subsection{The integrated CO depletion factor} \label{fd}  

  The CO integrated depletion factor is the ratio of the CO canonical abundance 
  ([CO]/[\HH]$\equiv 9.5 \times 10^{-5}$, \citeauthor{frerking1982} \citeyear{frerking1982}) 
  and the CO observed abundance.
  The CO observed abundance integrated along the line of sight can be calculated from the ratio of
  the CO column density and the \HH \ column density.
  \CEIO \ is measured to be optically thin in some of the cores (L1521F, L183, L694-2, TMC2, L492)
  where \CSEO \ observations are available.  
  Optical depth was derived from the integrated intensity ratio of the two isotopologues
  assuming equal excitation temperature and the local interstellar medium relative abundance value 
  \citep[3.2, ][]{wilson1994}. The maximum optical depth was measured to be 0.8 towards L1521F.
  Thus, for practical purposes, we assumed \CEIO \ optically thin in all the cores.
  To derive $N$(CO), we used the CTEX approximation on the \CEIO \ lines. 
  As already shown in \citet{caselli2002b} and \citet{crapsi2004}, this approach reduces the evaluation 
  of the integrated CO depletion factor to the simple formula 
  f$_{\rm D} =  0.085 \cdot {\cal F}_{\rm 1.2mm}/W_{\rm C^{18}O}$,
  where ${\cal F}_{{\rm 1.2mm}}$ is the 1.2-mm observed brightness in mJy~beam$^{-1}$ and 
  $W_{\rm C^{18}O}$ is the \CEIO \ integrated intensity in \kks.

  The sources of uncertainties in  this technique are several, and it is important to keep them in mind in
  the analysis of the results. 
  First, the  ``canonical CO abundance'' measured by different authors or in different objects varies  by a 
  factor of 2 \citep{lacy1994,alves1999,kramer1999}.
  Second, one has to bear in mind that the integrated CO depletion factor is an average value along the 
  line of sight, thus the local depletion factor in the nucleus of the core is much greater 
  \citep[see, e.g., ][]{caselli2002b,crapsi2004}.

  The integrated depletion factors of L1689B, L328, OphD and L429 were already 
  published in \citet{bacmann2002}, that of L1544 in \citet{caselli2002b} and that of L1521F in \citet{crapsi2004}. 
  We evaluate the f$_{\rm D}$  values for L1495, L1498 and L1517B using the data published in \citet{tafalla2002}, 
  for B68 using the \CEIO(1--0) intensities published in \citet{bergin2002} and the 1.2-mm continuum data published 
  by \citet{bianchi2003},  for L183 using the 1.2-mm map of \citet{pagani2003} and our \CEIO \ observations, 
  for L694-2 using the 1.2-mm map of M. Tafalla et al. (2005, in preparation) and our \CEIO \ observations and for 
  TMC2 and L492 using  the \CEIO \ and  the 1.2-mm maps from this work.
  Results are shown in column (4) of Table \ref{Tfd_nh2}.
  
\begin{deluxetable}{lccccccc}
\tablewidth{0pc}
\tabletypesize{\footnotesize}
\tablecaption{\HH \ densities, integrated CO depletion factor, flattened radius, aspect ratios and \NTHP(1--0) skewness. \label{Tfd_nh2} }
\tablehead{
core       & $N$(\HH)          &     $n$(\HH)	   & f$_{\rm D}$(CO)\tablenotemark{a}	  &  $r_{70}$     & aspect ratio  & aspect ratio  & \NTHP(1--0)  \\
           & $10^{22}$~\persc &   $10^{5}$~\percc   &  		  &  $10^{3}$ AU  & 70\% contour  & 50\% contour      & skewness \tablenotemark{b}
 }
\startdata

   L1498  &    3.2$\pm$1.0  &	1.0$\pm$0.7  &    7.5$\pm$2.5  &  6.6$\pm$0.7  &   1.6$\pm$0.3  &   1.5$\pm$0.3 &  -0.01$\pm$0.05\\
   L1495  &    3.1$\pm$1.0  &	1.1$\pm$0.7  &      7$\pm$2.4  &  4.7$\pm$0.6  &   2.0$\pm$0.4  &   1.9$\pm$0.4 &   0.10$\pm$0.05\\
  L1521F  &   13.5$\pm$2.2  &  11.0$\pm$1.8  &     15$\pm$3.6  &  3.4$\pm$0.5  &   1.2$\pm$0.1  &   1.6$\pm$0.2 &   0.19$\pm$0.06\\
    TMC2  &    6.0$\pm$1.2  &	3.0$\pm$0.8  &     13$\pm$3.3  &  5.2$\pm$0.6  &   1.2$\pm$0.4  &   1.1$\pm$0.3 &  -0.01$\pm$0.05\\
  L1517B  &    3.7$\pm$1.0  &	2.2$\pm$0.8  &    9.5$\pm$2.8  &  4.0$\pm$0.5  &   1.4$\pm$0.1  &   1.2$\pm$0.1 &  -0.08$\pm$0.05\\
   L1544  &    9.4$\pm$1.6  &  14.0$\pm$2.2  &     14$\pm$3.4  &  3.2$\pm$0.4  &   1.6$\pm$0.2  &   1.9$\pm$0.2 &   0.40$\pm$0.09\\
    L183  &   10.0$\pm$1.7  &  10.0$\pm$1.7  &     12$\pm$3.1  &  4.8$\pm$0.6  &   1.8$\pm$0.2  &   2.5$\pm$0.2 &   0.09$\pm$0.05\\
    OphD  &    8.2$\pm$1.5  &	8.5$\pm$1.5  &     14$\pm$3.4  &  6.1$\pm$0.7  &   2.1$\pm$0.4  &   1.9$\pm$0.4 &   0.54$\pm$0.12\\
     B68  &    1.4$\pm$0.3  &   0.8$\pm$0.7  &    3.4$\pm$2.1  &  3.7$\pm$0.9  &   1.3$\pm$0.3  &   1.3$\pm$0.3 &  -0.09$\pm$0.05\\ 
    L492  &    4.4$\pm$1.1  &	2.1$\pm$0.8  &      8$\pm$2.6  &  7.8$\pm$0.8  &   1.5$\pm$0.2  &   2.0$\pm$0.2 &   0.25$\pm$0.07\\
    L328  &    5.7$\pm$1.2  &	1.8$\pm$0.7  &    8.5$\pm$2.6  &  3.7$\pm$0.5  &   1.1$\pm$0.2  &   1.2$\pm$0.2 &  -0.02$\pm$0.08\\
    L429  &    8.8$\pm$1.6  &	6.0$\pm$1.1  &   15.5$\pm$3.7  &  3.6$\pm$0.5  &   1.4$\pm$0.2  &   2.1$\pm$0.3 &  -0.20$\pm$0.10\\
  L694-2  &    7.8$\pm$1.4  &	9.0$\pm$1.5  &     11$\pm$3.0  &  5.5$\pm$0.6  &   1.3$\pm$0.1  &   1.4$\pm$0.1 &   0.22$\pm$0.07\\

\enddata
\tablecomments{For reference to the literature data see text.}		
\tablenotetext{a}{The CO depletion factor in L1689B is 4.5$\pm$2.2.}		
\tablenotetext{b}{\NTHP(1--0) skewness was measured  also in:  
L1495A-N  (-0.04$\pm$0.05);
L1400K    (-0.42$\pm$0.10);
TMC1	  (-0.13$\pm$0.06);
TMC1C	  (0.43$\pm$0.10);
L1512	  (0.02$\pm$0.05);
L234E-S   (-0.25$\pm$0.07);
L1197	  (0.08$\pm$0.11);
CB246	  (0.02$\pm$0.05).}
\end{deluxetable}

 \subsection{\HH \ volume density} \label{nh2}
 
  The \HH \ column density can be derived from the millimeter continuum under the 
  approximation of optically thin emission, constant temperature and constant emissivity
  of the dust at these wavelengths.
  The equation that relates the \HH \ column density to the 1.2-mm flux is:
 
 \begin{displaymath}
   N(H_2) = \frac{S_{1.2mm}}{B_{\nu}(T)\,\Omega_{beam} \; \kappa_{1.2mm} \; m } ,
 \end{displaymath}
  where $N$(\HH) is the \HH \ column density, S$_{1.2~mm}$ is the flux density integrated over the 
  solid angle $\Omega_{beam}$,  $B_{\nu}(T)$ is the Planck function at temperature $T$, $\kappa_{1.2mm}$ is the
  dust opacity per unit mass  and $m$ is the  mean molecular mass. In our calculations, we assume a dust 
  temperature of 10~K  and a dust opacity of 0.005~cm$^2$~g$^{-1}$. The assumed dust 
  temperature is justified by the estimates obtained from \AMM \  \citep{tafalla2002} and
  \HTCO \  \citep{bacmann2002} observations in several cores of our sample.
  We note that the dust temperature is predicted to drop in the core nucleus \citep{evans2001,zucconi2001}; 
  assuming a temperature of 8~K would cause an increase of the column density of a factor of 1.5.
  We also remark that the dust opacity value, assumed following \citet{andre1996}, suffers from an uncertainty 
  of a factor of 2 \citep{henning1995,bianchi2003,kramer2003}.

  In this work we both collected 1.2-mm maps from the literature and observed new sources with the
  bolometers.
  In Table~\ref{Tfd_nh2} we report the \HH \ volume density at the peak of L429 and L328 taken from \citet{bacmann2000}, 
  of  L1495,  L1498,  L1517B and L1544 from  \citet{tafalla2002}, of  OphD and L1689B from
  \citet{wardthompson1999}, of  L694-2 from \citet{harvey2003} (using their Bonnor-Ebert
  fit and changing $\kappa_{1.2mm}$ to 0.005~cm$^2$~g$^{-1}$ for consistency), of L183 from \citet{pagani2003}
  and of L1521F from \citet{crapsi2004}.
  All the central densities above were determined from 1.2-mm continuum data and using the same basic technique 
  (i.e., a fit of the observed continuum emission starting from a volume density model of the core).
  We remark that these estimates were obtained assuming spherical symmetry. A strong deviation of the density profile
  along the line of sight with respect to what is seen in the plane of the sky will cause errors in the volume 
  density determination. 
  For some of these cores other different  estimates  for the peak density are reported in the literature; these
  were obtained starting from different data (molecular line or 850~$\mu$m continuum) or using different techniques.
  In the case of multiple estimates all done starting from  1.2-mm continuum data, we adopted the most recent value, 
  e.g., L1544 has three determinations of the density that were done from a 1.2-mm map: 
   1.5$\times 10^6$~\percc \ by \citet{wardthompson1999}, 4$\times 10^5$~\percc \ by \citet{bacmann2000} 
  and 1.4$\times 10^6$~\percc \ by \citet{tafalla2002}; in this work we adopted the most recent determination 
  from \citet{tafalla2002}.
  
  The \HH \ central volume density at the TMC2, L492 and B68 peaks was evaluated with the same technique as for L1521F 
  \citep{crapsi2004}. We assumed the analytical model for the volume density given by \citet{tafalla2002}: 
  $n(H_2)=n_0/(1+(r/r_0)^\alpha)$, where the central density ($n_0$), the ``flattened radius'' ($r_0$) and the steepness
  of the profile at large radii ($\alpha$) are free parameters. Then, we searched for the best combination of parameters
  able to predict the observed continuum  measurements.  
  In this way we found at the peak of TMC2 an \HH \ density equal to $3 \times 10^5$~\percc, while 
  $2.1 \times 10^5$~\percc \ is the measured density towards  the L492 peak. The B68 central density, derived from 
  lower  S/N data (and with the worst angular resolution), was found to be $8 \times 10^4$~\percc \ with large error bars.

  Volume densities were also derived using an LVG program (see Section~\ref{LVG}), whose
  results are summarized in Table~\ref{TLVG}.
  We found systematically lower central densities using the molecular data ($\sim 3$ times less) than using the dust
  \citep[as already found in L1544; see][]{caselli2002b}, suggesting that \NTHP \ and \NTDP \ may be probing a shell 
  exterior to the high-density nucleus (see Figure~\ref{Fnh2lvg} in 
  Section~\ref{LVG}). This indication is consistent with the idea that N$_2$ could freeze--out for densities higher
  than $5 \times 10^5$~\percc \ as suggested by  \citet{bergin2002} and  \citet{belloche2004}. 
  \citet{keto2004}  found a similar result using
  \NTHP  \ and suggest that a way to reconcile gas density estimates from
  dust observations with those from molecular line observations is to increase
  the dust mass opacity to $ \kappa = 0.04$~cm$^2$~g$^{-1}$, a value typical of
  fluffy aggregates with ice mantles.
  We stress that higher resolution observations are strongly needed to confirm this trend.

  \subsection{Dust emission equivalent radius and aspect ratio} \label{r70}
 
  In starless cores the \HH \ density profile can be approximated by a region 
  of roughly constant density  followed by a power-law decrease 
  \citep{wardthompson1999,bacmann2000,tafalla2002}; this behaviour is reflected in the 1.2-mm continuum emission. 
  Since the ``flattened nucleus'' is expected to  become smaller with the ongoing of infall, its size 
  gives us a measure of the relative contraction stage  reached by the cores. 
  Unfortunately each author uses different definitions for the size of the flattened region; thus, we have measured,
  for the 13 cores with a 1.2-mm map, the  area  within the  70\% contour of the dust peak. 
  We then translated this into a radius (reported as $r_{70}$ in figures and tables) by evaluating the square root of this 
  area divided by $\pi$.
  The resulting values are reported in column (5) of Table~\ref{Tfd_nh2}.
  This approach is useful for a relative comparison of the different cores. 
  The choice of the 70\% contour as threshold is justified by the fact that we wanted to limit the influence of the
  ``cometary tail'' (see Section~\ref{shape}) on the $r_{70}$ determination.
  
  We also evaluated the aspect ratio of the dust emission measuring the major and minor axis size of the 50\% contour 
  and 70\% contour (results are shown in columns (6) and (7) of Table~\ref{Tfd_nh2}).
  We found that the axis ratio increases at lower contours in the more cometary cores, reaching values up to 2.5.

 \subsection{The line asymmetry} \label{skew}
 
  The analysis of the line profile gives us clues on the core kinematics.
  In fact, motions along the line of sight combined with optical depth effects, molecular 
  abundance variations and temperature gradients alter the line profile, causing departures from a Gaussian line
  shape.   In particular, looking at the density peak, we expect a ``blue shoulder'' in the case of outflow motions 
  and a ``red shoulder'' in the case of infall motions \citep[see ][for a review]{evans2002}.
  
  Here, we determine the degree of line asymmetry by calculating the line
  skewness, or the third moment, of a distribution equivalent to the line profile.
  This is defined as $\sum_{i=1}^{N}{((x_i-\bar{x})/\sigma_{x})^3}/N$, where $\bar{x}$ and $\sigma_{x}$ are the 
  first and second moments of the same distribution and $N$ is the total number of points 
  \citep[see e.g.][]{press1992}. 
  By definition, the third moment measures the deviation from a symmetric distribution, as the first moment furnishes 
  its mean and the second  yields its dispersion. It is a non-dimensional quantity and assumes positive
  values for distributions that have a ``tail'' in the values higher than the mean and negative values 
  in the opposite case.
  We applied this analysis to the ($F_1 \, F = 2 \, 2 \rightarrow 1 \, 1$) component of  \NTHP \ 
  (the fourth starting from the lower velocities).
  The ``isolated'' ($F_1 \, F = 0 \, 1 \rightarrow 1 \, 2$, the one at lowest velocity) and the ``weak'' 
  ($F_1 \, F = 1 \, 0 \rightarrow 1 \, 1$, the one at highest velocity) components  show approximately 
  the same behaviour as this one in the very high S/N spectra (RMS $< 0.03$~K with
  a 0.063~\kms \ channel spacing); thus, in order to enlarge the
  sample to spectra with slightly lower S/N (RMS $< 0.1$~K), we used the data from the 
  ($F_1 \, F = 2 \, 2 \rightarrow 1 \, 1$) component, which also provides a sufficient isolation from the other
  components even in the spectra with broadest line widths. Moreover, given the role of the optical depth in the
  production of the asymmetric profiles, the ($F_1 \, F = 2 \, 2 \rightarrow 1 \, 1$) component is supposed
  to have a more pronounced skewness than the other two components.
  Results are shown in Table~\ref{Tfd_nh2}.
  Eight cores show red-skewed ``contracting'' profiles, six cores show blue-skewed ``expanding'' profiles, and seven cores have
  symmetric profiles within the error bars. Note that the line peak in the ``contracting'' profiles falls at lower 
  velocities (or on the blue side) than $\bar{x}$ whereas asymmetric profiles with outflow character are the reverse.

\section{Discussion} \label{dis}
In the following we discuss the relations between the quantities determined in the previous 
sections and compare them with simple theoretical expectations.
We  divide this section into two main parts: in the first we concentrate on the connection between
deuterium fractionation and the other chemical and dynamical parameters derived in Section~\ref{ana}, 
while in the second we cross-correlate only the parameters linked with the kinematical activity.

\subsection{Deuterium Fractionation correlations}  
\subsubsection{Deuterium fractionation and CO depletion} \label{dfr_fd}
  
  An observational proof of the chemical model relating the deuterium fractionation with the CO depletion factor
  was given in \citet{bacmann2003} using the ratio of D$_2$CO over H$_2$CO in a sample of five cores. 
  Although D$_2$CO and H$_2$CO have been shown to deplete in
  the inner nucleus of the pre--stellar cores \citep{carey1998,maret2004} and hence  do not trace the
  very inner nucleus, these five cores (included  in our sample) showed a good correlation between the CO depletion 
  and the deuterium fractionation in  H$_2$CO with the exception of OphD.
  
  Another possible test is to check for a correlation between the CO depletion factor and the
  deuterium fractionation within a given core. This test was performed in \citet{crapsi2004} showing again
  an affirmative answer.
  
\begin{figure}[!tbp]
 \begin{center}
 \resizebox{8cm}{!}{\includegraphics{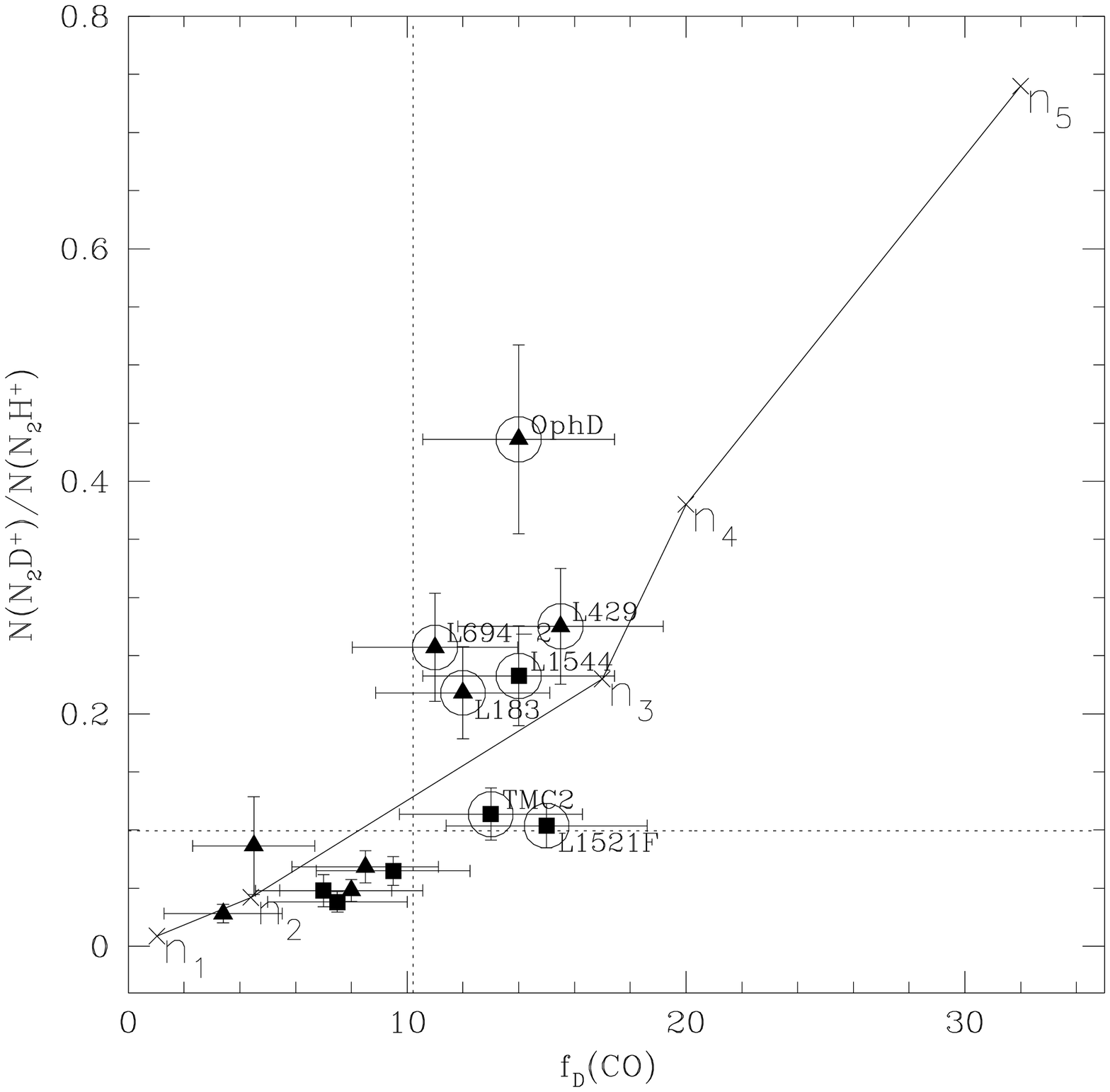}}
 \caption{Deuterium fractionation vs. integrated CO depletion factor at the \NTDP \ peak position. Depletion factor  values
	  were taken from the literature and from present data (see Section \ref{fd} for references).
	  The solid line connects the predictions of chemical models of spherically symmetric dense cores with 
	  various degrees of central concentrations (from central densities of 4.4$\times 10^4$~\percc \ in model 
          $n_1$ to 4.4$\times 10^8$~\percc \  in model $n_5$), roughly simulating (in spherical symmetry) the 
	  evolutionary sequence of contracting disk-like clouds in the \citet{ciolek2000} model (see text). 
	  Dotted lines are the arithmetic means of the distributions. Cores belonging to the Taurus Molecular 
	  Cloud are marked by a square; pre--stellar cores candidates are circled.
	 \label{Fdfrfd}}
 \end{center}
\end{figure}

  Here we searched for a correlation between the integrated CO depletion factor and the deuterium
  fractionation in a subsample of 14 cores using the high density gas tracers \NTHP \ and \NTDP \ to derive the 
  deuterium fractionation. The result is shown in Figure~\ref{Fdfrfd}, where we denote by enclosed data
  points our "candidate pre--stellar cores", defined for the purpose of this work as having \DoH \ $\ge 0.1$. 
  
  The correlation is not extremely tight (correlation coefficient 69\%; 74\% if only Taurus cores are considered), 
  but indeed we can say that the cores that show higher CO depletions do also show higher deuterium fractionation.
  As in \citet{bacmann2003}, OphD seems to have ``too much'' deuterium fractionation compared to the CO depletion. 

  In Figure~\ref{Fdfrfd} we also show theoretical curves from simple chemical models
  based on that described by \citet{caselli2002b}. These models all
  assume that the gas and dust temperatures are constant across the core
  and equal to 10 K.  The chemical network and chemical parameters are also
  the same in all the models: apart from \HH , the chemical network
  contains the three neutral species CO, \MOLN , and O, which can freeze--out
  onto dust grains and return to the gas phase via thermal desorption or
   cosmic--ray impulsive heating \citep[following the formulation of ][]{hasegawa1993}.
  On the other hand, the abundances of the molecular ions (\NTHP , \HCOP , \HTHOP ,  \HTHP \ 
  and their (multiply) deuterated counterparts ) are calculated in terms of the
  instantaneous abundances of the neutral species.  This simplification is 
  based on the fact that the ``ion chemistry'' timescale is much shorter than the 
  depletion timescale (see \citealt{caselli2002b} and 
  \citealt{crapsi2004}for 
  details). Here we included the multiply deuterated forms of \HTHP , without
  any distinction between the ortho and para forms of molecular hydrogen and 
  \HTWDP \  and without taking into account the so--called back-reactions 
  between, e.g., ortho--\HH \ and ortho--\HTWDP \  \citep{gerlich2002}, which limit 
  the deuterium fractionation, as discussed in \citet{walmsley2004}. 
  However, we adopted the new value of the rate coefficient for the proton--deuteron exchange 
  reaction \HTHP + HD $\rightarrow$ \HTWDP + \HH \  \citep[3.5$\times$10$^{-10}$~cm$^3$~s$^{-1}$;][]{gerlich2002}.  
  The other parameters used are: the cosmic--ray ionization
  rate $\zeta$ = 1.3$\times$10$^{-17}$~s$^{-1}$; the CO, \MOLN, and O  binding 
  energies: $E({\rm CO})$ = 1210 K, $E({\rm N_2})$ = 800 K, and $E({\rm O})$ = 650 K
  \footnote{All the adopted parameters are the same as in \citet{crapsi2004}, 
  with the exception of $E({\rm O})$, which is now 100 K lower than before.
  The decrease in this parameter is necessary to keep the deuterium fractionation
  at a similar level as in \citet{crapsi2004}, for the particular case of L1521F,
  after the inclusion of all the multiply deuterated forms of \HTHP . See also 
  \citet{caselli2002b} for the importance of atomic oxygen in this simple 
  chemical network.} respectively; and the lower cutoff radius of the MRN \citep{mathis1977} 
  distribution $a_{\rm min}$ = 5$\times$10$^{-6}$ cm.

  The different model predictions, denoted as $n_1$, $n_2$, $n_3$, $n_4$ and  $n_5$ in Figure~\ref{Fdfrfd},
  refer to model spherical clouds with different density structures, in
  (rough) analogy with the density structure of the model cloud undergoing
  infall, described by \citet[hereafter CB00]{ciolek2000}, at successive
  evolutionary stages (from $t_1$ = 2.27~Myr to $t_5$ = 2.684~Myr:
  see CB00). We remark that our calculations assume the steady state and thus our
  model makes use of the density profile only, and not of the time scale.
  We also note that the CB00 model has cylindrical (rather than
  spherical) symmetry and we have neglected this fact assuming  the same
  dependence on spherical radius as that on cylindrical radius in CB00.
  We assume that the density profiles are given by the
  analytical formula of \citet{tafalla2002}, with different values of the
  parameters (namely, the central density, the radius of the inner flat region,
  and the asymptotic power index) to approximately reproduce the density profiles
  of Figure~1a of CB00.   In particular, the central densities 
  of the cores at each time are given by 10$^j \times 4.37 \times 10 ^3$~\percc \ 
  (for $j$ = 1, ..., 5) as in CB00. The abundance profiles 
  obtained in this way have then been converted into column densities via 
  integration along the line of sight and successive convolution with a 
  two-dimensional Gaussian, simulating observations with an HPBW of 22\arcsec . 
  From Figure~\ref{Fdfrfd} we note that the deuterium fractionation in \NTHP \ and 
  the observed CO depletion factor ($f_{\rm D}$) are predicted to increase with core 
  evolution and actually explain the observed trend, although some scatter
  is expected in the data, given the different environments where they are 
  immersed and their different formation histories.  
  In particular, we expect magnetic field strength, amount of turbulence, 
  external radiation field and external pressure to influence the evolution of a core.
  Hence two cores with the same mass and same age could have different degrees of evolution
  depending on the characteristics of the surrounding star-forming region. 
   We note that most of the ``pre--stellar'' cores are found
  close to the $n_3$ result, which, in the CB00 model, is the model of best
  agreement with the measured density profile of L1544. Therefore, we
  can conclude that the "pre--stellar core" condition, \DoH \ $\ge 0.1$, 
  is consistent with values of central densities $n$(\HH) $\ge 10^6 $~\percc, 
  in agreement with  \HH \ central  densities measured from the dust in Section~\ref{nh2}
  for these cores.

  We also searched for correlations between the depletion factor and the \HH \ column density 
  and found a positive  correlation (87\% correlation coefficient; 90\% for Taurus only), confirming the general 
  chemical scenario about CO depletion \citep[see][]{dalgarno1984,roberts2000a,bacmann2002}.
  Several authors have investigated the chemical structure of pre--stellar
  cores, coupling together dynamical and chemical evolution with
  detailed models \citep{bergin1997,aikawa2001,li2002,shematovich2003}. 
  Deuterium and singly deuterated species were,
  however, only included in \citet{aikawa2001}, who were able to reasonably reproduce the observed molecular
  D/H ratios toward L1544, although the column density of \NTHP \ was
  underestimated.  \citet{aikawa2003} overcame this problem by
  adding surface chemistry to their models. Surface processes significantly
  enhanced the production of the parent molecule \MOLN , increasing the
  \NTHP \ column density to the observed values.

   More recently, \citet{roberts2004} studied the chemistry of
   pre--stellar cores including multiply deuterated species but neglecting
   the dynamical evolution. For the particular cases of L1544 and OphD
   they found \DoH \ column density ratios significantly (factor
   of about 5) larger than the observed values, suggesting that the
   deuterium fractionation process is probably too efficient in their models.  Possible
   causes of this disagreement are: (i) the exclusion of the so--called
   back reactions of deuterated isotopologues of \HTHP \ with ortho-\HH \  \citep{gerlich2002}
    which may lower
   the molecular D/H ratios \citep{walmsley2004}; (ii) the use of large
  (radius of 0.1~$\mu$m) dust grains in the chemical network, which
   underestimates the recombination of molecular ions compared to models
   where a population of smaller grains is present or where an MRN size
   distribution of dust particles is considered; (iii) the use of UMIST
   rates, which produce \HTHP \ and analogues more efficiently than the "New
   Standard Model" from  Ohio State University \citep[see ][for details]{roberts2004}.

\subsubsection{Deuterium fractionation vs. \HH \ central density, dust equivalent radius, \NTHP \ line width and  \NTHP \ line skewness} \label{dfr_kin}

  In theoretical simulations, we expect an increase of the central $n$(\HH) density and a decrease of
  $r_{70}$, the equivalent radius of the flat region, with ongoing contraction 
  \citep[e.g.,][]{lizano1989,foster1993,ciolek1995,li1999}.
  Those two parameters can be considered as  indicators of ``dynamical evolution'';
  here we cross-correlate them with the deuterium fractionation.  
  
\begin{figure}[htbp]
 \begin{center}
 \resizebox{14.9cm}{!}{\includegraphics{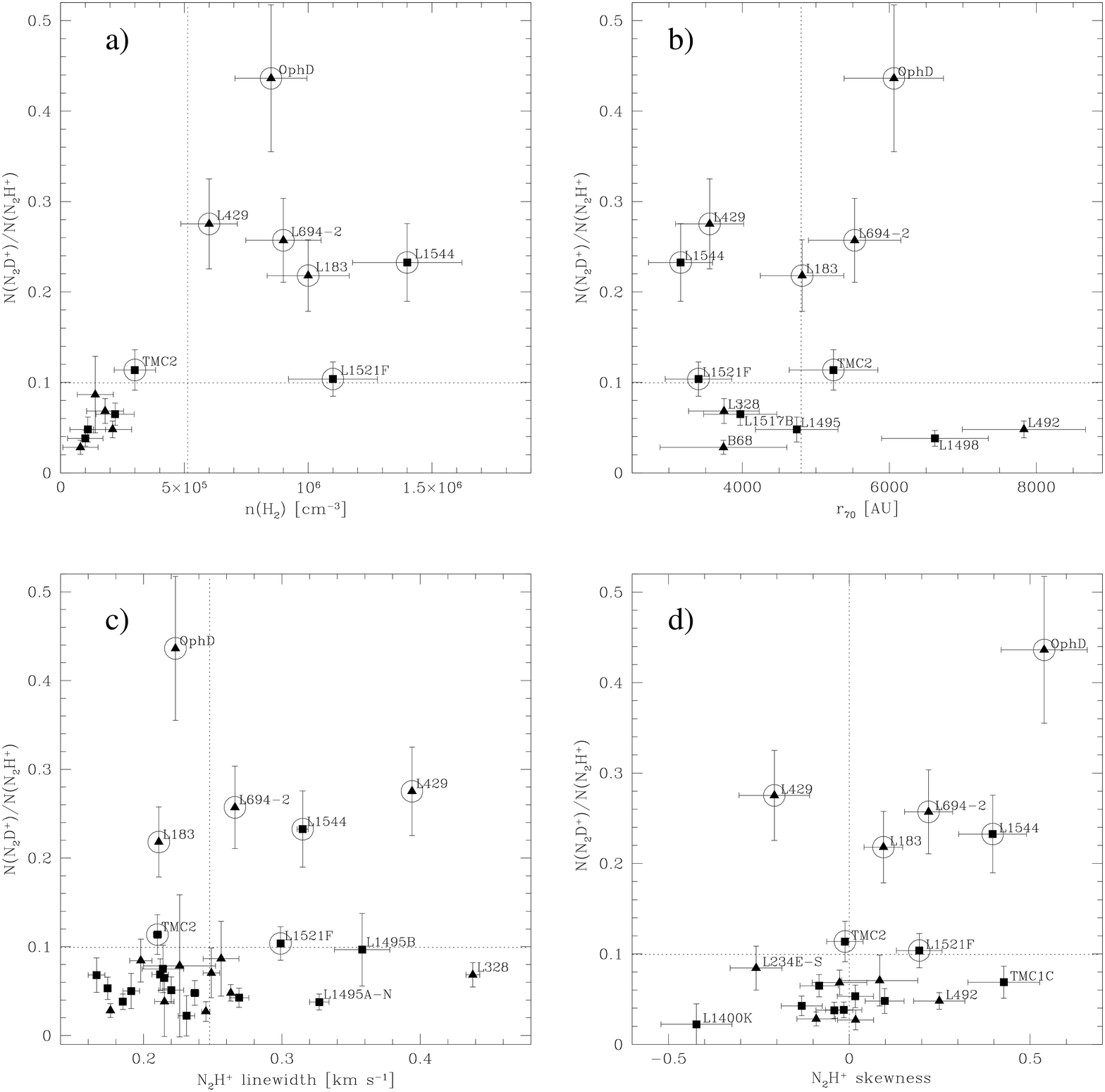}}
 \caption{ Deuterium fractionation correlations. In all panels: dotted lines are the arithmetic means of the distributions (except in panel d)); 
         cores belonging to the Taurus Molecular Cloud are marked by a square; 
	 pre--stellar cores candidates are circled; the values were evaluated at \NTDP \ peak position.
       {\bf Panel a):} Deuterium fractionation  vs. \HH \ central volume  density.  The $n$(\HH) values are 
	 taken from the literature (see Section~\ref{nh2}), with the exception of TMC2, L492 and B68, estimated in this work.
       {\bf Panel b):} Deuterium fractionation  vs. equivalent radius of the flattened region 
	 $r_{70}$ is defined as $\sqrt{area_{70}/\pi}$, with $area_{70}$ as the area within the 70\% contour in the 1.2-mm map.
       {\bf Panel c):} Deuterium fractionation vs. \NTHP (1--0) line width evaluated from the simultaneous fit of all the 7 hyperfine 
         components. 
       {\bf Panel d):} Deuterium fractionation vs. \NTHP (1--0) line asymmetry. The vertical dotted line separates
         the red-skewed spectra from the blue-skewed ones (the arithmetic mean is 0.07).        
	  \label{Fdfr}}
 \end{center}
\end{figure}

  From Figure~\ref{Fdfr} panel a) one can see that there is a positive correlation between
  \DoH \ and $n$(\HH) apart from OphD and L1521F which show opposite behaviours from the rest of the cores.
  Six of our pre--stellar cores candidates have \HH \ volume densities above the average (5.1$\times 10^5$~\percc),
  and all of these show increases in deuterium fractionation above the average.
  Our  estimate of the volume density depends on the  assumed spherical geometry: hence, we also considered 
  the correlation between the deuterium fractionation and the \HH \ column density. While the general 
  trend remained the same, the peculiarity of OphD and L1521F  becomes more marked.\\ 

  We check here the relative variation of the flattened region in the 13 cores for which we have a
  1.2-mm map. Having treated all the cores in a consistent way allows us to consider the differences 
  in $r_{70}$ real  and not due to differences in the adopted fitting model.\\
  The correlation between deuterium fractionation and $r_{70}$ is shown in Figure~\ref{Fdfr} panel b).
  A close relationship of the flattened radius with the deuterium fractionation is not found.\\
  OphD shows the biggest variation from this picture, but one should notice the presence of a second peak
  inside its 70\% contour. It is thus possible that the OphD equivalent radius is overestimated. A similar 
  argument holds for L183.
  Limiting the sample to the Taurus cores significantly tightens the correlation, increasing the 
  correlation coefficient  from 1\% to 64\%.  Once again, this result suggests that cores embedded in 
  different environments (e.g. external radiation field) hardly represent a homogeneous sample in which to study   
  evolutionary trends.\\

  \citet{caselli2002a} showed that in L1544 there was a definite trend for the 
  \NTHP \ and \NTDP \ line widths to increase towards  the high-density region.
  This result was interpreted as a sign of increasing infall activity towards the center.
  The same trend, although less marked, is seen in L1521F \citep{crapsi2004}.\\      
  In this fashion, we take here the \NTHP \ line width as an indicator of infall activity and thus proximity 
  to the critical state for dynamical collapse and
  we check if the cores that had larger line widths are the same as those with the largest \DoH \ ratio, 
  to search for links between  physical and chemical evolution.

\onecolumn
\twocolumn

  \noindent   The result is in Figure~\ref{Fdfr} panel~c).
  In our sample we do not see any correlation for larger line width in the nucleus of cores with larger 
  deuterium fractionation.
 In particular, we note that L328 shows broad lines but low \DoH \ and that OphD  has the
  opposite behaviour.  
  The two broadest line cores, L328 and L429, have spectra with both flat tops  and 
  blue-skewed asymmetry (see Sect. \ref{skew}), suggesting that their lines could be broadened for reasons
  other than infall. 
  We attempted to avoid the environmental differences, as a source of scatter for the correlation, limiting 
  the sample to the Taurus cores only; in this case 
  the correlation coefficient increases from 22\% to 40\%. \\
  We point out that the above correlation does not make any distinction 
  between cores that do and do not show evidence of central infall (through the \NTHP (1--0) line
  width broadening toward the dust peak position, or with the presence of 
  ``red shoulders''), so that the different line widths observed in the cores
  of the present sample may simply arise from different amounts of turbulence 
  in their interiors or other types of motions (e.g., expansion or oscillations) 
  along the line of sight. \\
  
  We also try to correlate the line skewness with the \DoH \ ratio in Figure~\ref{Fdfr} panel~d).
  Also in this case there is a general agreement with the proposed idea that cores with higher deuterium
  fractionation also show signs of kinematical evolution. The main exceptions in this case are given by 
  L429 on the one hand and  TMC1C and L492 on the other.
  It is interesting to note that the correlation between \DoH \ and the line asymmetry
  is more tight  than that with the line width.

\subsubsection{Deuterium fractionation and extended infall} \label{dfr_cwl}

  \citet{lee1999, lee2001} and \citet{lee2004} in a series of papers conducted a search for infall motions towards a
  sample of 70 starless cores.
  Infall motions were identified through the presence of velocity shifts between high density tracers (\NTHP(1--0) probing 
  densities up to a few times $10^6$~\percc)  and low density tracers (CS(2--1) tracing the gas up to $\sim 10^4$~\percc)
  and through the observation of double-peaked self-absorbed optically thick lines  whose blue peak was brighter than 
  the red peak \citep{lee2004}.
  Given the CS depletion in the inner core, those measurements gauge the infall of the external envelope only.
  
  Twenty-four cores in our sample were observed also in the Lee et al. papers (\citeyear{lee2001, lee2004}), 
  including our chemically 
  evolved candidates L1544, L1521F, L694-2, L183, L429, TMC2 and OphD (there known as L1696A).
  Considering now only this common subsample, \citet{lee2004} found that 12 of these cores
  show infall signatures in at least three of the seven tests they performed on CS, \DCOP \ and \NTHP \
  lines, thus they are considered strong infall candidates.
  All our chemically evolved candidates (\DoH $> 0.1$) 
  \citep[except OphD, which was not studied with all the techniques by][]{lee2004}
  were identified as strong infall candidates by the \citet{lee2004} analysis. 
  This seems to show that our chemically evolved candidates, the majority of which show clear signs of 
  central infall, are a subsample of the cores undergoing extended contraction.
  
  We also performed a direct comparison between one of the infall indicators in \citet{lee2004}, $\delta$V$_{\rm DCO^+}$ 
  (${\rm \equiv (V_{\rm DCO^+}-V_{\rm N_2H^+})/\Delta V_{\rm N_2H^+}}$), and our \NTHP(1--0) skewness
  on a common subsample of 15 cores. We find a very good agreement between those two parameters with only two cases 
 in which they give a different indication: L429 which has significant outflow signature in the \NTHP(1--0) skewness
  but marginal infall indication in $\delta$V$_{\rm DCO^+}$,  and L183, which has a very strong outflow behaviour in 
   $\delta$V$_{\rm DCO^+}$ whereas it shows an infall profile in our  \NTHP(1--0) spectrum.
  For the 13 remaining cores the two indicators correlate with a 70\% correlation coefficient.

\subsection{Dynamical parameters correlations} 

  We cross-correlated  our ``kinematic activity'' indicators, i.e., the \NTHP (1--0) line width, the 
  radius of the flattened region, the \HH \ column density, and the \NTHP (1--0) line asymmetry, to see whether 
  they give  consistent information.

\onecolumn
\begin{figure}[htbp]
 \begin{center}
 \resizebox{14.9cm}{!}{\includegraphics{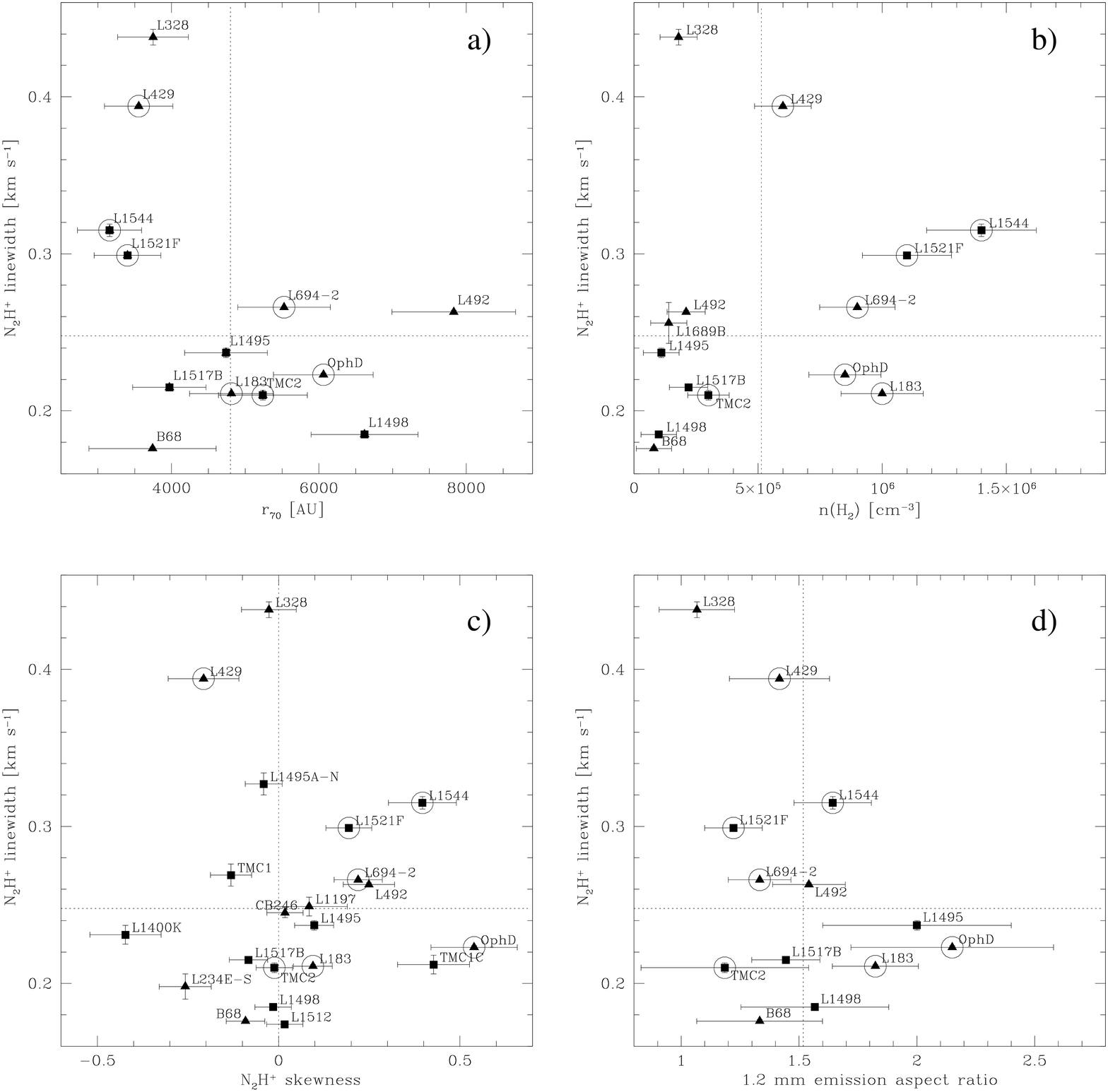}}
 \caption{Correlations between ``dynamical evolution'' indicators. In all panels: dotted lines are the arithmetic means of the distributions 
         (but in panel c));  cores belonging to the Taurus Molecular Cloud are marked by a square; 
	 pre--stellar cores candidates are circled; the values were evaluated at \NTDP \ peak position.
      {\bf Panel a):} \NTHP (1--0) line width vs. equivalent radius of the flattened region 
	 $r_{70}$ is defined as $\sqrt{area_{70}/\pi}$, with $area_{70}$ as the area within the 70\% contour in the 1.2-mm map.
      {\bf Panel b):} \NTHP (1--0) line width vs. \HH \ volume density. The $n$(\HH) values are 
	 taken from the literature (see Section~\ref{nh2}), with the exception of TMC2, L492 and B68, estimated in this work.
      {\bf Panel c):} \NTHP (1--0) line width vs. \NTHP (1--0) line asymmetry. The vertical dotted line here separates
         the red-skewed spectra from the blue-skewed ones (the arithmetic mean is 0.07). 
      {\bf Panel d):} \NTHP (1--0) line width vs. aspect ratio of the 50\% contour of the 1.2-mm emission.	
      \label{Fdv}}
 \end{center}
\end{figure}
\twocolumn

  \noindent The result is encouraging. We can in fact see in panel~a) of Figure~\ref{Fdv}  that $r_{70}$ is anti-correlated with 
  \NTHP(1--0) line width (correlation coefficient 67\%; 86\% limiting to Taurus cores); 
  note that  even cores that in the other plots do not behave like the rest of the sample 
  (OphD and L328) lie here in the common trend.\\
  
  In panel~b) of Figure~\ref{Fdv} we find that, with the notable exceptions of L328 and L429,
  the \NTHP (1--0) line width correlates  fairly well with the \HH \ central density (correlation coefficient 20\% for the
  total sample; 64\% without L328 and L429; 94\% for only the Taurus cores).

  In the picture given in Section~\ref{dfr_kin}, the cores with strong motions along the line of sight should 
  show stronger asymmetries and broader lines towards the core peak. In Figure~\ref{Fdv} panel~c) 
  we plotted the relation between the width and the skewness of the \NTHP \ line. 
  Although again a tight correlation is not found, we can see that there is a general trend of increased
  skewness in  broader spectra. Remarkable exceptions are constituted by OphD and TMC1-C, having narrow lines
  strongly skewed towards the red, and by L429 and L328, which have the broadest lines in the sample 
  but do not show any infall asymmetry.  Also in this case, limiting the sample 
  to Taurus cores, the correlation coefficient increases  from 2\% to 21\%.

\subsubsection{Line width and aspect ratio} \label{dvasprat}

  The fact that the majority of the cores in our sample (see Figure~\ref{Fcomet} in Section~\ref{shape}) show large 
  departures from spherical 
  symmetry has considerable significance for our understanding of the dynamics of these objects.
  Core shape could  for example reflect the magnetic field structure and indeed  several of the cores in our sample 
  show linear polarization of around 5\%-10\%  \citep[e.g.,][]{crutcher2004}, suggesting magnetic fields of order 
  100~$\mu$G.  However, the observed structures could also be a consequence of dynamical interactions
  with the core surroundings as is suggested by the observed ``cometary'' shapes.
  It is natural therefore to search for  kinematic evidence in our \NTHP \ maps  that either excludes or 
  favors one of these scenarios.

  There are various possible kinematic probes, including the observed velocity field, but the simplest is to check for
  a correlation between line width and aspect ratio.  Here the interpretation depends considerably on whether the 
  true core shape is ``close to oblate'' or ``close to prolate'' 
  \citep[see, e.g., ][who have attempted to decide this on the basis of the statistics of observed aspect ratios]{jones2002,curry2002}.  
  Present results are inconclusive and we conclude that one must consider both possibilities.

  In either case, if cores are magnetically dominated, motion along the field lines will be facilitated. 
  We would expect that in such cases, motions will predominantly be parallel to the axis  of symmetry  and hence 
  for oblate cores, one might expect the largest line widths for face-on circular sources and for
  prolate cores for end--on  (also circular) objects.
  On the other hand, ambipolar diffusion necessarily occurs across field lines and this will have the opposite
  effect. Thus, for example, edge-on oblate cores will have the largest line width. In fact, CB00 
  have provided predictions for how the velocity field due to the latter process might appear in L1544 and 
  \citet{caselli2002a} showed that the observations were roughly consistent with the model for this object.

  However, this does not seem to be the case in L1521F \citep{crapsi2004} and also not in a more general sample 
  of cores as shown in Figure~\ref{Fdv} panel~d), where we plot the \NTHP(1--0) line width against aspect ratio from the millimeter 
  continuum maps. There is no clear trend in this plot, and clearly a larger sample  is needed. However, our results
  do not give evidence that in general there are increased line widths towards elongated core nuclei as one might expect  
  on the basis of models such as those of CB00.  On the other hand, our results also do not seem to 
  favor the models of \citet{curry2001} which suggest that initially elongated structures should become more
  spherical as they evolve owing to ambipolar diffusion.  We conclude that it would be useful to have more detailed predictions
  for model pre--stellar cores of the evolution with time of both the line width and the velocity field.

\subsection{Associated properties} \label{summ}
\begin{deluxetable}{llcccccccccccccc}
\tablewidth{0pc}
\tabletypesize{\footnotesize}
\tablecaption{Chemical and kinematical evolution ``probes'' in the 13 cores with full data set (\NTHP, \NTDP, 
\CEIO \ and 1.2-mm emission). \label{Tsumm}}
\tablehead{
  core	 & \rotatebox{90}{L1544} & \rotatebox{90}{L1521F} & \rotatebox{90}{L694-2} & \rotatebox{90}{L429} & \rotatebox{90}{L183} 
  & \rotatebox{90}{OphD}  & \rotatebox{90}{TMC2} & \rotatebox{90}{L492} & \rotatebox{90}{L328}  & \rotatebox{90}{L1495} & \rotatebox{90}{L1517B} 
  & \rotatebox{90}{B68} & \rotatebox{90}{L1498} 
 }
\startdata
  $N$(\NTDP) $>1.0 \times 10^{12} $ \persc  &  1   &   1   &  1   &  1	&  1   &  1   &  1   &   0  &  0  &   0   &  0  &  0  &   0  \\
  $N$(\NTHP) $>8.5 \times 10^{12} $ \persc  &  1   &   1   &  1   &  1	&  1   &  0   &  1   &   1  &  1  &   1   &  1  &  1  &   0  \\
  $\frac{[N_2D^+]}{[N_2H^+]} > 0.10$        &  1   &   1   &  1   &  1	&  1   &  1   &  1   &   0  &  0  &   0   &  0  &  0  &   0  \\
  f$_{\rm D}$(CO) $> 10.2$ 	 	    &  1   &   1   &  1   &  1	&  1   &  1   &  1   &   0  &  0  &   0   &  0  &  0  &   0  \\
  $n$(\HH) $> 5.1 \times 10^{5} $ \percc    &  1   &   1   &  1   &  1	&  1   &  1   &  0   &   0  &  0  &   0   &  0  &  0  &   0  \\
  $\Delta V_{N_2H^+}$ $> 0.25$ \kms         &  1   &   1   &  1   &  1	&  0   &  0   &  0   &   1  &  1  &   0   &  0  &  0  &   0  \\
  infall asym. ($skewness > 0$) 	    &  1   &   1   &  1   &  0	&  1   &  1   &  0   &   1  &  0  &   1   &  0  &  0  &   0  \\
  $r_{70}$ $< 4800$ AU                      &  1   &   1   &  0   &  1	&  1   &  0   &  0   &   0  &  1  &   1   &  1  &  1  &   0  \\
 \hline
  total  		                    &  8   &   8   &  7   &  7	&  7   &  5   &  4   &   3  &  3  &   3   &  2  &  2  &   0  \\
 	     
\enddata								          
\tablecomments{ For each value, but the skewness, the thresholds  are given by the arithmetic mean value for the sample.}
\end{deluxetable}

  We have shown in this section a series of properties that we think are associated with
  an evolved stage of  starless cores either because of enhanced chemical activity or because of an
  advanced kinematical process.
  
  Even though we did not find tight one-to-one correlations between the numerical values it appears
  that these properties globally isolate a group of cores that look peculiar with respect to the rest of the sample.
  We summarize these characteristics of the evolved starless cores, or pre--stellar cores, in  Table~\ref{Tsumm}.
  L1544 and L1521F were found to be more evolved than the ``average'' core in each test we performed, while
  L183, L429 and L694-2 failed just one test and can be considered strong candidates for pre--stellar cores
  as well. Other cases, such as OphD and TMC2, are more doubtful.

  In the presence of an unbiased survey and assuming that all the cores will eventually form a star, 
  one could attempt to estimate the duration for the pre--stellar core phase 
  by dividing the number of pre--stellar core candidates by the total number of objects in the sample.  
  Since our main selection criterion stated that only the sources with stronger \NTHP(1--0) observations 
  (in the \citealt{lee2001}sample) or stronger dust emission-absorption (in the \citealt{bacmann2000} sample)
  were observed, we cannot consider our survey unbiased.
  Nevertheless, taking the number of \NTHP--detected cores within 250~pc (49) in the \citet{lee1999} sample 
  as  representative of  the total number of the \NTHP--bright cores and   
  supposing that L1544, L1521F, L183, L429, L694-2, OphD and TMC2  are the only pre--stellar cores observable 
  in this sample, we can derive a duration of the pre--stellar phase equal to
  7/49, or $\sim$14\% of the lifetime of a core since it started to be observable in \NTHP.
  Taking the smallest central density observed in our sample (B68: $n$(\HH) $\sim 8\times 10^4$~\percc) as 
  the threshold for \NTHP \ emission, we can derive a lifetime of an \NTHP--bright core of $\sim$0.4~Myr 
  (using the CB00 ambipolar diffusion models); hence, the pre--stellar core phase should last around
  60,000~yr. 
  An alternative estimate can be made evaluating the free-fall time for the threshold density for a pre--stellar core 
  ($n$(\HH)$\sim 5.1 \times 10^5$) resulting in a duration of $\sim$60,000~yr.\\
  The pre--stellar core lifetime can also  be estimated  from the comparison with the 
  estimated time spent as a  Class 0 protostar \citep[$\sim 2 \times 10^5$~yr][]{visser2002}.
  From  \citet{andre2000} and \citet{visser2002}, we derive a total number of 11 Class~0 protostars within 250~pc.
  This, compared to the seven evolved cores found in this work, gives a duration of the pre--stellar phase of $\sim$120,000~yr, 
  longer than the estimates above. However, we feel that this discrepancy merely reflects the great 
  uncertainties in both estimates.

\section{Conclusions} \label{con}
  
  We collected a set of  \NTHP, \NTDP, \CEIO \ and 1.2-mm continuum observations using the IRAM~30-m
  for a sample of 31 starless cores, completing  literature data with new observations.
  We retrieved from these data chemical and kinematical parameters supposed to discriminate 
  young from more evolved cores such as: deuterium fractionation, integrated CO depletion factor, 
  \HH \ density, \NTHP \ line width and line asymmetry.
  Our conclusions are summarized as follows:

  1. We attempted several correlations between the chemical evolution indicators (deuterium fractionation, integrated CO 
  depletion factor and \NTHP \ and \NTDP \ column density) and properties of dynamically evolved cores (such as \HH \ volume 
  density, line width and asymmetry and size of the flattened region).
  Although we did not find very tight dependencies between these quantities, we recognized that all these properties, as
  a whole more than taken singularly, draw from the total sample those cores with an enhanced state of evolution.
  As can be seen from Table~\ref{Tsumm}, we find that, compared to the total sample, L1544, L1521F, L183, L429, 
  L694-2, OphD and TMC2 globally show:
   \begin{itemize}
    \item higher \NTHP \ column density
    \item higher \NTDP \ column density
    \item higher \DoH \ abundance ratio
    \item higher \HH  \ column density
    \item higher integrated CO depletion factor
    \item broader \NTHP \ lines
    \item smaller flattened radius
    \item stronger infall asymmetries
   \end{itemize}
   with some exceptions in each category.
   
   2. Limiting the sample to the cores belonging to the Taurus Molecular Cloud tightens the 
   correlations, probably by diminishing possible effects due to environmental differences such as
   magnetic field strength, amount of turbulence, external radiation field and external pressure.

   3. The picture emerging from the observations is consistent with
   predictions of chemical models accounting for molecular freezeout,
   where cores with higher central densities show a lower fraction of CO
   in the gas phase and a correspondingly larger deuterium enrichment in
   the core nucleus.

   4.  \NTDP \ and \NTHP \ emissions, unlike \CEIO  , peak very close to the dust continuum maxima in the mapped cores, 
   confirming the fact that those species probe better the conditions of the high density gas.
   Moreover, the maps of \NTDP \ and  \NTHP(3--2) were found to be systematically smaller than the \NTHP(1--0) ones
   in all the mapped cores,  consistent with the idea  that these trace better the high-density nucleus.
   We note that central \HH \ densities evaluated from the dust emission were found systematically larger than
   those calculated from LVG modeling of \NTHP \  and \NTDP \ data, suggesting that depletion of \MOLN \ could possibly
   occur at the highest density peak, although finer resolution data are needed to confirm this effect.

   5. The decrease of line-width towards the edge observed in L1544 \citep{caselli2002a} and L1521F 
   \citep{crapsi2004} has been found in only one other core (L694-2) with supposedly the same degree of evolution.

\acknowledgements

We would like to thank the IRAM 30-m staff for help during the observations.
We gratefully acknowledge  Philippe Andr\' e, Aurore Bacmann, Simone Bianchi and Laurent Pagani for providing us
with electronic versions of their 1.2-mm data.  
We also thank Luca Dore for communicating new frequency determinations. 
A. C. was partly supported by NASA ``Origins of Solar System Grant'' (NAG 5-13050).
P. C. and C. M. W. acknowledge support from the MIUR grant ``Dust and molecules in astrophysical environments''. 
C. W. L. acknowledges supports from the Basic Research Program (KOSEF R01-2003-000-10513-0) of the Korea 
Science and Engineering Foundation.

\appendix
\section{Maps} \label{maps}
   
   Figures \ref{Fl694-2} to  \ref{Ftmc2} show  maps towards L183, OphD, L429, TMC2, L492 and L694-2, which, together with 
   L1544 and L1521F, stand out from the rest of the sample as having very bright \NTDP (2--1)
   lines. Similar maps for L1544 can be found in \citet{caselli2002a} and \citet{tafalla2002} and for L1521F in 
   \citet{crapsi2004}.
   The \NTHP \ and \NTDP \ data shown in Figures~\ref{Fl694-2}-\ref{Fl429} were all convolved with a 26\arcsec \ beam in order to 
   reach the same spatial resolution as the \NTHP (1--0) data and make the comparison easier. The \NTHP(1--0) map of TMC2 
   (Figure~\ref{Ftmc2}) was kept with the original FCRAO resolution (54\arcsec), while we have not smoothed  the \NTDP(2--1) map
    of L492  (Figure~\ref{Fl492}) because of the small extension of it.
   Similarly, all  the 1.2-mm maps  were smoothed to a 22\arcsec \ resolution, the beamsize of the IRAM~30-m at the \CEIO(1--0) frequencies.
   
   The most evident feature of Figures \ref{Fl694-2} to  \ref{Ftmc2} is that the \CEIO \ does not trace the 
   dust continuum while the \NTHP \ and \NTDP \ emission morphology is similar to that of the dust, as already shown by 
   several authors \citep[e.g. ][]{caselli1999,bergin2001,tafalla2002} .
   While this is true for L1544, L1521F, L429, L694-2 and OphD, there are differences 
   between the 1.2-mm map and the \NTHP--\NTDP \ maps for L183 and TMC2.
   For example, the L183 dust emission peaks 30\arcsec \ to the south of \NTHP(1--0) and, especially, \NTDP(2--1).
   
   TMC2 shows an even more complex situation; in fact, the 1.2-mm maps revealed much more structure than the \NTHP(1--0)
   map of \citet{caselli2002c} led us to believe. Whereas we found a local 1.2-mm peak next to the \NTHP(1--0) one, we also found
   brighter peaks of the dust emission in regions where the \NTHP(1--0) is fainter. 
   Note that the \NTHP \ map in Figure~\ref{Ftmc2} has a lower angular resolution (HPBW = 54\arcsec) than in the other sources;
   smoothing the dust continuum map to this resolution does not help to reconcile the two emissions, so we present 
   here the 1.2-mm map with the same resolution as in the other cores (i.e. 22\arcsec).
   We also add in Figure~\ref{Ftmc2} the position of a few radio stars and of a couple of $IRAS$ point sources; the presence 
   of these objects could explain the discrepancy between the \NTHP \ and the dust maps.

   Another common feature clearly seen throughout Figures \ref{Fl694-2} to  \ref{Ftmc2} is that
   the \NTDP \ maps are  systematically smaller than the \NTHP \ ones (see Table~\ref{Tgau2d}). This could indicate that the deuterated species 
   are better tracers of the core nucleus.
   Similar behaviour is seen when the higher transition maps are compared with those at lower $J$; 
   this may be caused by the increase in density towards map center.
\begin{deluxetable}{lccccc}
\tablewidth{0pc}
\tabletypesize{\footnotesize}
\tablecaption{\NTHP, \NTDP \ and 1.2-mm maps full width at half maximum.\label{Tgau2d} }
\tablehead{ core           & 1.2mm   & \NTHP(1--0)  & \NTHP(3--2)  &   \NTDP(2--1)    &   \NTDP(3--2)        }
\startdata
       L183    & 185\arcsec$\times$ 67\arcsec  & 217\arcsec$\times$ 92\arcsec  &    \nodata                 & 93\arcsec$\times$ 72\arcsec  &     \nodata                  \\
        OphD    & 129\arcsec$\times$ 70\arcsec  & 170\arcsec$\times$ 70\arcsec  &    \nodata                 & 93\arcsec$\times$ 62\arcsec  &  67\arcsec$\times$ 65\arcsec \\
        L492    & 180\arcsec$\times$100\arcsec  &  98\arcsec$\times$ 78\arcsec  &    \nodata                 &      \nodata                 &    \nodata                   \\
        L429    & 102\arcsec$\times$ 53\arcsec  &  91\arcsec$\times$ 78\arcsec  & 72\arcsec$\times$53\arcsec & 60\arcsec$\times$ 56\arcsec  &  43\arcsec$\times$ 41\arcsec \\
      L694-2    &  77\arcsec$\times$ 55\arcsec  &  85\arcsec$\times$ 67\arcsec  &    \nodata                 & 68\arcsec$\times$ 60\arcsec  &     \nodata                  \\
\enddata								          
\end{deluxetable}	
   
   In OphD (see Figure~\ref{Fophd}), we note the presence of two nuclei embedded in the same envelope; both have a similar brightness
   in the 1.2-mm continuum although they have a very different intensity in the \NTHP(1--0), \NTDP(2--1) and  \NTDP(3--2)
   (the \NTHP(3--2) has poor S/N towards the northeast  core). 
   The relative intensity of the two nuclei decreases going from the lowest density tracer (\NTHP(1--0)) to the highest density
   one (\NTDP(3--2)), suggesting that the southwest core is more centrally peaked than the northeast one.

\begin{figure}[!htbp]
 \begin{center}
 \caption{L694-2 maps  in 1.2-mm continuum, \CEIO(1--0), \NTHP (1--0) and \NTDP (2--1) . 
 \NTHP  \ and \NTDP \ data were smoothed to a common resolution of 26\arcsec \ and regridded; 
 the reference position is the one reported in Table~\ref{Tsample}.
 The 1.2-mm map (M. Tafalla et al. 2005, in preparation) was smoothed to a resolution of 22\arcsec .
 Contour levels are 30\% to 90\% by 15\% of the peak for each map. Peak values are
 185~mJy/22\arcsec~beam for the 1.2-mm map and 5.16, 1.26 and 1.76 \kks \ for \NTHP (1--0), 
 \NTDP (2--1) and \CEIO(1--0) respectively. Available 40\arcsec$\times$40\arcsec \ maps of \NTHP \ and \NTDP(3--2)
 are not reported because of poor S/N outside the peak.
 \label{Fl694-2}}
 \end{center}
\end{figure}

\begin{figure}[!htbp]
 \begin{center}
 \caption{L183 maps in 1.2-mm continuum, \CEIO(1--0), \NTHP (1--0)  and \NTDP (2--1). Same as in 
 Figure~\ref{Fl694-2} but for the peak values: here 239~mJy/22\arcsec~beam for the 1.2-mm map and
 4.31, 1.65, and 2.5 \kks \ for \NTHP (1--0), \NTDP (2--1) and \CEIO(1--0) 
 respectively. The 1.2-mm map has been taken from \citet{pagani2003}. Available 30\arcsec$\times$60\arcsec \
 maps of \NTHP and \NTDP(3--2) are not reported because of poor S/N outside the peak.
 \label{Fl183}}
 \end{center}
\end{figure}

\begin{figure}[!htbp]
 \begin{center}
 \caption{OphD maps in 1.2-mm continuum, \NTHP (1--0) and (3--2) and \NTDP (2--1) and (3--2). Same  as in 
 Figure~\ref{Fl694-2} but for the peak values: here 195~mJy/22\arcsec~beam for the 1.2-mm map and
 3.81, 0.49, 1.75 and 0.60  \kks \ for \NTHP (1--0) and (3--2), \NTDP (2--1) and (3--2) respectively. 
  The 1.2-mm map has been taken from \citet{wardthompson1999}. The RMS of the \NTHP (3--2) increases 
  significantly towards north east, thus contours above -20\arcsec \ have to be taken with caution.
  \label{Fophd}}
 \end{center}
\end{figure}

\begin{figure}[!htbp]
 \begin{center}
 \caption{L429 maps in 1.2-mm continuum, \NTHP (1--0) and (3--2) and \NTDP (2--1) and (3--2).  Same  as in  
 Figure~\ref{Fl694-2} but for the peak values: here 210~mJy/22\arcsec~beam for the 1.2-mm map and
 6.12, 0.63, 2.07and 0.69 \kks \ for \NTHP (1--0) and (3--2), \NTDP (2--1) and (3--2) respectively. 
 The 1.2-mm map has been taken from \citet{bacmann2000}. \label{Fl429}}
 \end{center}
\end{figure}

\begin{figure}[!htbp]
 \begin{center}
 \caption{L492 maps in 1.2-mm continuum, \CEIO(1--0), \NTHP (1--0)  and \NTDP (2--1).  Same  as in  
 Figure~\ref{Fl694-2} but for the peak values: here 114~mJy/22\arcsec \ beam for the 1.2-mm map and
 3.96, 0.31 and 1.69 \kks \ for \NTHP (1--0), \NTDP (2--1) and \CEIO(1--0) 
respectively.  \label{Fl492}}
 \end{center}
\end{figure}

\begin{figure}[!htbp]
 \begin{center}
 \caption{TMC2 maps in 1.2-mm emission smoothed to 22\arcsec \ (contours in both panels) and \CEIO(1--0) (gray scale in the top 
 panel) from new 30m observations and the \NTHP(1--0) map (gray scale in the bottom panel) 
 taken by  \citet{caselli2002c} with FCRAO (HPBW 54\arcsec; reference position here is the same as in Table \ref{Tsample}). 
 Contour levels starts from 30\% of peak and increase by 15\%
 of it. Peak values are: 138~mJy/22\arcsec~beam for the 1.2-mm map, 2.12 \kks \ for the \CEIO \ observations.
 The \NTHP(1--0) intensity peak is 3.6 \kks \ in $T_{MB}$ temperature. 
 In the bottom panel the positions of seven young stars  (in order of ascending RA:  JH93, JH90, JH91, JH92, JH94 - 
 \citeauthor{wendker1995} \citeyear{wendker1995} ; FY Tau, FZ Tau - \citeauthor{nurmanova1983} \citeyear{nurmanova1983}) 
are denoted by stars. Two $IRAS$ point sources visible in the same field are denoted by a filled square. 
 \label{Ftmc2}}
 \end{center}
\end{figure}

\section{Core shapes and kinematics} \label{shape}

  We underline a marked tendency of the starless cores for having the dust emission peak shifted relative to the
  centroid of the cores, suggesting that the cores either have been 
  compressed by external agents or have moved relatively to the 
  surrounding environment \citep[but see ][]{walsh2004}. 
  In Figure~\ref{Fcomet} we attempt a classification of the shape of the starless cores evaluating a parameter able to 
  quantify the degree of ``cometariness'' or how much the dust peak is shifted relative to the lower contours of the map.
  We evaluated the centroid of the map as the mean position of the points within the 50\% contour and 
  defined the degree of ``cometariness'' as the distance between this and the dust peak in units of $r_{70}$, the equivalent radius 
  of the core. 
  OphD and TMC2 were not included in this classification since they harbour multiple peaks within their 50\% contour 
  (see Figures~\ref{Fophd} and  \ref{Ftmc2}); thus, the ``cometariness'' parameter is meaningless.
  From this small subsample we do not notice any clear correspondence between a more marked cometary shape and other kinematical
  properties of the core (e.g. \HH \ density, \NTHP(1--0) line-width, or $r_{70}$).\\

\begin{figure}[!htbp]
 \begin{center}
  \caption{The 1.2-mm continuum maps of 12 cores in our sample. Light contours are spaced from 10\% to 90\% by 10\% of the peak,
  while the 98\% and 50\% contours are represented by a thick line. The cross marks the mean position of the points
  within the 50\% contour. The cores are ranked for increasing ``cometariness''. This parameter is reported at the top right corner of
  each map and is calculated as the distance between the mean point and peak in units of $r_{70}$.
  TMC2 and OphD were not included in this classification since they have multiple peaks within their 50\% contour, so this 
  technique failed for them. \label{Fcomet}}
 \end{center}
\end{figure}

\begin{figure}[htbp]
 \begin{center}
 \caption{\NTHP(1--0) velocity gradients towards the cores L694-2, OphD, L183, L429 and L492.
 The \NTHP (1--0) data were regridded in a regular pattern.  The integrated intensity maps are shown in greyscale 
 (see Figures \ref{Fl694-2} - \ref{Fl492}).  
 Local velocity gradients in the adjacent nine points are represented by an arrow pointing in the direction of increasing velocity
 and with the length proportional to the magnitude (1~\kms~pc$^{-1}$ corresponds to a 6\arcsec-long arrow). Note that the angular 
 scale of each map is the same. \label{Fvgrad}}
 \end{center}
\end{figure}

\begin{figure}[!htbp]
 \begin{center}
 \resizebox{12cm}{!}{\includegraphics{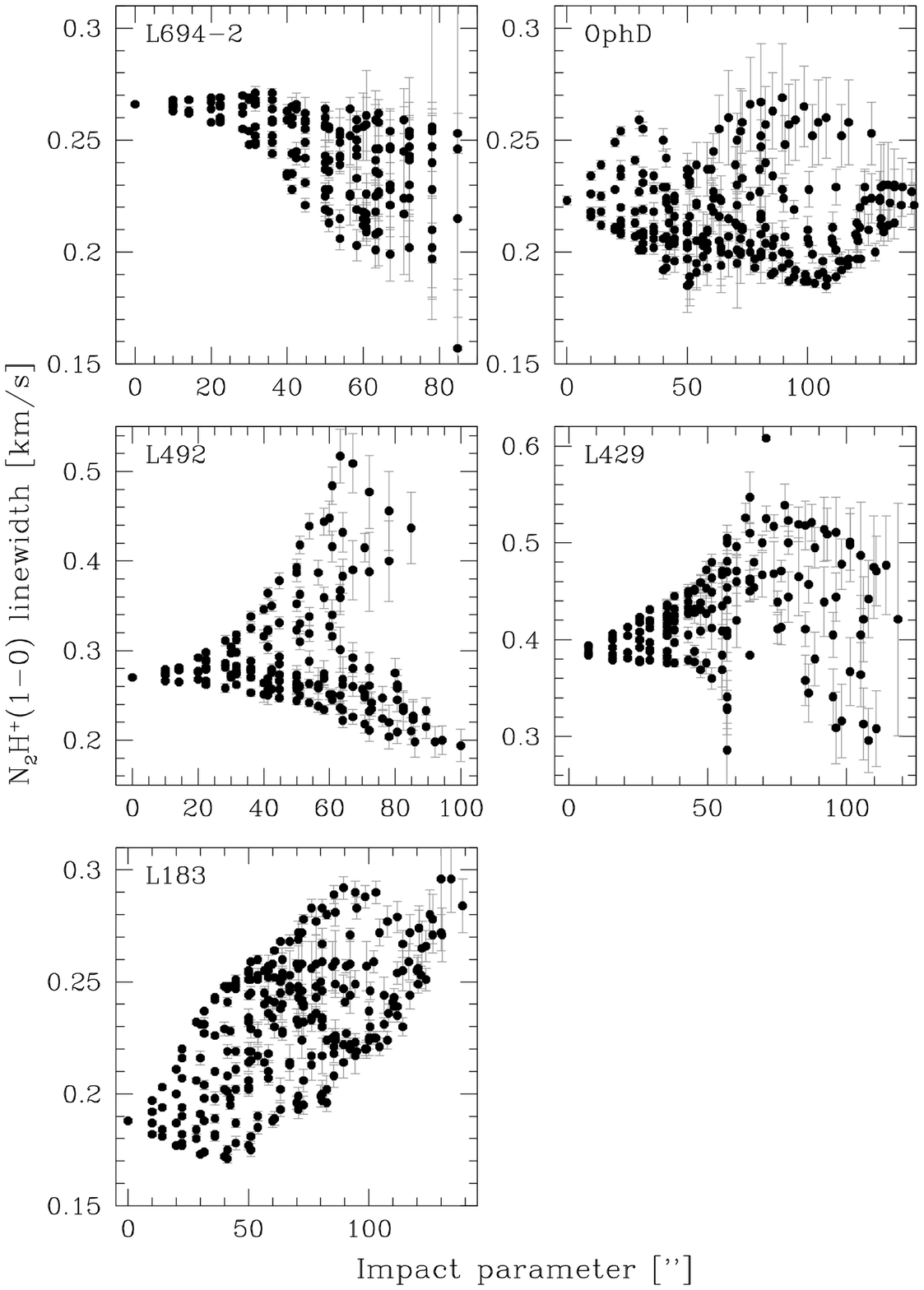}}
 \caption{\NTHP(1--0) intrinsic line width gradients within the cores L694-2, OphD, L183, L429 and L492. 
 The angular offset was evaluated from the \NTDP(2--1) peak position (see Table \ref{Tinten}). \label{Fdvgrad}}
 \end{center}
\end{figure}

  Using the \NTHP(1--0) velocities from the hyperfine structure fitting, we found strong velocity variations in L183 
  (0.23~\kms \ in a 0.16~pc length) and  OphD (0.19~\kms in a 0.12~pc length), while  a smaller variation 
  (0.07~\kms \ in a 0.1~pc length)  was found in L694-2 and  constant velocity fields
  were seen in the inner 0.08~pc of L429 and L492. 
  We evaluated the total velocity gradients  of these cores under the assumption of solid body rotation
  \citep[see ][]{goodman1993}. The magnitudes and the direction of the velocity gradients  
  are reported in columns (1) and (2) of Table~\ref{Tvg}. 
\begin{deluxetable}{lccc}
\tablewidth{0pc}
\tabletypesize{\footnotesize}
\tablecaption{Global and local velocity gradients from \NTHP(1--0) maps. \label{Tvg} }
\tablehead{
core       & ${\cal G}$     & $\Theta ^{a}$ & $<{\cal G}_{\rm l}> ^{b}$  \\
           & \kms~pc$^{-1}$ & \arcdeg       &  \kms~pc$^{-1}$               
	   }
\startdata
    L183   & 1.4$\pm$0.1    &   -49$\pm$1   &  1.5$\pm$0.7   \\
    OphD   & 1.6$\pm$0.1    &  -119$\pm$1   &  1.6$\pm$1.0   \\
  L694-2   & 0.7$\pm$0.1    &   -18$\pm$1   &  0.8$\pm$0.3   \\
    L429   & 0.4$\pm$0.1    &   -33$\pm$2   &  1.9$\pm$1.1   \\
    L492   & 0.2$\pm$0.1    &    83$\pm$2   &  1.5$\pm$0.8   \\

\enddata
\tablenotetext{a}{Direction of increasing velocity, measured East of North.}		
\tablenotetext{b}{Mean values of the magnitude of local gradients and corresponding standard error.}
\end{deluxetable}	
  These calculations agree with the velocity variations seen across the cores.
  \\
  Local velocity gradients, evaluated in  $3 \times 3$ adjacent points, were also calculated towards these cores 
  \citep[see ][ for details on the procedure]{caselli2002a}. The analysis was performed on regridded data and 
  results are shown in Figure~\ref{Fvgrad} and Table~\ref{Tvg}. 
  L183 and OphD present  strong motions and a regular velocity field.
  L429 and L492 show very little motion in the inner parts but larger gradients at the edge of the 
  map with arrows pointing in different directions on different sides of the cores, explaining the absence of velocity
  variation in the inner part and the small total velocity gradient.
  Finally, L694-2 shows the smallest local gradients with a regular velocity field only in the inner part.

  Using the \NTHP(1--0) and \NTDP(2--1) maps, we also checked for the presence of  line width gradients in L694-2, OphD, 
  L183, L429 and L492  in the same fashion as for L1544 \citep{caselli2002a}, B68 \citep{lada2003} and L1521F 
  \citep{crapsi2004}.
  From figure \ref{Fdvgrad} we find that only L694-2 shows the line width decrease that was evidenced in L1521F and L1544, 
  while a marginal similarity is found in the first 50\arcsec \ around OphD and in the northern part of L492.
  The opposite behaviour is seen in L429 and L183 (as found in B68).
  The \NTDP(2--1) line widths evidenced the same behaviour as in figure \ref{Fdvgrad}.

\section{Density determination in the LVG approximation} \label{LVG}

  The Large  Velocity Gradient (LVG) approximation simplifies the radiative transfer treatment assuming that a photon
  produced in a contracting or expanding cloud is reabsorbed locally and that the local velocity gradient is much larger than the
  ratio of line width to cloud size.  In addition, one normally assumes homogeneous and isothermal conditions. 
  Assuming a temperatutre of 10~K, we have used this approximation, with collisional rates from
  \citet{green1975} to compute level populations for \NTHP \ and \NTDP \ and to predict line intensities for these species as a
  function of density and species column density. 
  We then compared the model predictions with the observed \NTHP(3--2)/\NTHP(1--0) intensity ratio and the
  \NTHP(1--0)  integrated intensity (or the \NTDP(3--2)/\NTDP(2--1) and \NTDP(2--1) intensity) exploiting the fact that
  the former is mainly sensitive to $n$(\HH) and the latter to the \NTHP (\NTDP) column density.
  Our results are extremely sensitive to the  intensity of the two 3--2 transitions and we hence confine our discussion to the
  eight cores with S/N$> 5$ in the 1~mm lines.
  An extra constraint comes from the total $\tau$ of \NTHP(1--0) (\NTDP(2--1)). Note that, given the small opacities 
  found in \NTDP(2--1) (and in many \NTHP(1--0) spectra), $\tau$ has little influence on the $n$(\HH) determination whereas
  it is a useful constraint for column density.

\begin{figure}[!tbp]
 \begin{center}
 \resizebox{11cm}{!}{\includegraphics{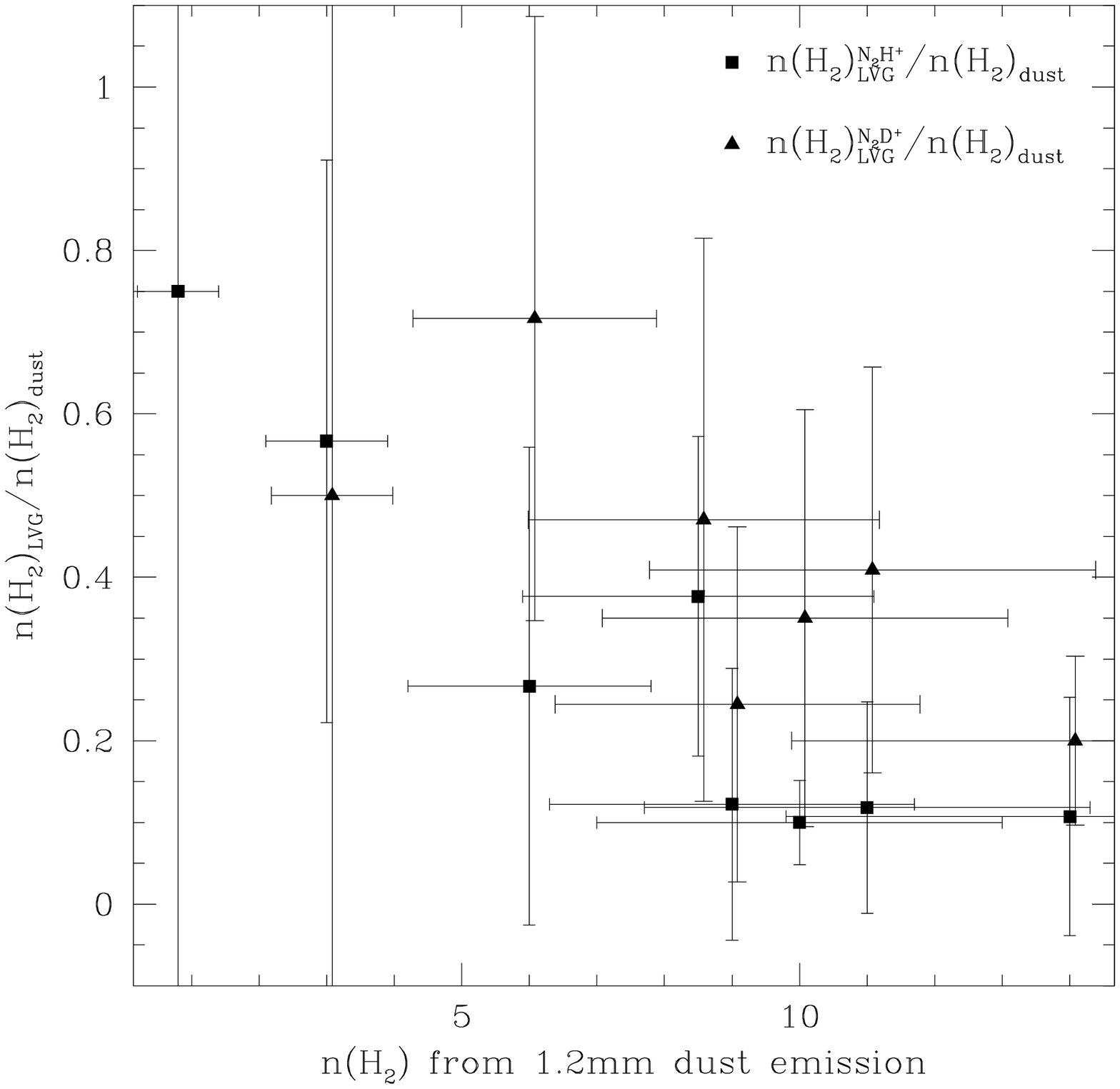}}
 \caption{\HH \ central density derived from dust observation compared with the determination from \NTHP/\NTDP \ observations.
 The density inferred from the dust emission is always higher than the one derived from the molecular data, although
 the difference seems to diminish in the lowest density cores. Moreover, the estimates done starting from \NTDP \ data
 furnish moderately higher \HH \ density, consistently with the idea that \NTDP \ is a better tracer of the dust emission.
 Data points from \NTDP \ were slightly offset in the $x$-axis in order to show them better.   
  \label{Fnh2lvg}}
 \end{center}
\end{figure}

  In Table~\ref{TLVG} we summarize the results of the LVG calculations.
  The ratio of the \NTHP to \NTDP \ column densities inferred from LVG and the relative value found in the CTEX approximation
  have an arithmetic mean value of 0.9 and a dispersion of  25\%. Thus, the LVG results  are consistent with the 
  column densities derived assuming a constant rotational temperature. 
  It is interesting to note that the \HH \ densities derived from the molecular data are  systematically
  smaller (a factor of $\sim 3$) than those  derived from the dust observations (see Figure~\ref{Fnh2lvg}).

\begin{deluxetable}{l|cc|cc}
\tablewidth{0pc}
\tabletypesize{\footnotesize}
\tablecaption{\HH \ volume density and \NTHP--\NTDP \ column densities derived in the LVG approximation. 
 \label{TLVG} }
\tablehead{
            &            \multicolumn{2}{c|}{\NTHP}   &	     \multicolumn{2}{c}{\NTDP}	     \\
core        & $n$(\HH)          &     $N$(\NTHP)	      & $n$(\HH)           &     $N$(\NTDP)	     \\
            & $10^{5}$\percc  &  $10^{13}$\persc      & $10^{5}$\percc   &  $10^{12}$\persc     	   }
\startdata
L1521F      &  1.3  &	 3.4   &    4.5  &  1.6  \\
TMC2        &  1.7  &	 1.0   &    1.5  &  1.5  \\
L1544       &  1.5  &	 2.8   &    2.8  &  4.6  \\
L183        &  1.0  &	 2.0   &    3.5  &  2.8  \\
OphD        &  3.2  &	 0.6   &    4.0  &  2.6  \\
B68         &  0.6  &    1.2   &\nodata &\nodata \\
L429        &  1.6  &	 3.1   &    4.3  &  4.9  \\
L694-2      &  1.1  &	 2.1   &    2.2  &  3.3  \\
\enddata								          
\tablecomments{Errors in the LVG determinations are estimated around 30\% from comparisons with CTEX determiantions.}
\end{deluxetable}	

  \noindent The differences between the two determinations diminish in the  low-density cores (a factor of $\sim 2$).
  These differences may partially be due to our assumptions concerning the dust millimeter opacity  and temperature 
  (see \citeauthor{galli2002}  \citeyear{galli2002} for a discussion of the latter and 
  \citeauthor{bianchi2003} \citeyear{bianchi2003} for a discussion of the former). 
  Another possibility is that depletion of N$_{2}$ (and hence of \NTHP \  and  \NTDP) may  occur in 
  regions with density higher than $5 \times 10^5$~\percc \ 
  \citep[see also ][]{caselli1999, bergin2001, caselli2002b, crapsi2004, belloche2004}. In the latter case, one would expect 
  differences between maps of \NTHP \ and dust continuum (one possible example could be L183, see Section~\ref{maps}).
  We note also that there is a hint in our data  that densities derived from \NTDP \ are  systematically higher than 
  those from \NTHP  \ and this again may be consistent with a hole caused by N$_{2}$  freeze--out. However, 
  we stress that these trends need confirmation  and that  measurements with higher sensitivity and angular 
  resolution are needed.

\end{document}